\documentclass[useAMS,usenatbib]{mn2e}

\usepackage{amsmath}
\usepackage{amssymb}
\usepackage[parfill]{parskip}   
\usepackage{ifpdf}
\usepackage{graphicx} 
\usepackage{longtable}
\usepackage{rotating}
\usepackage{epsf}
\usepackage{bm}
\usepackage{subfigure}
\usepackage{morefloats}
\usepackage{gensymb}
\usepackage{verbatim}
\usepackage{array}
\usepackage{lscape}

\graphicspath{{figures/}}

\begin{document} 

\title[Kinematics of dwarf galaxies]{Kinematics of dwarf galaxies in gas-rich groups, and the survival and detectability of tidal dwarf galaxies.}
\author[S. Sweet et al.]{Sarah M. Sweet$^{1*}$, Michael J. Drinkwater$^{2}$, Gerhardt Meurer$^{3,4}$, Virginia Kilborn$^{5}$,
\newauthor Fiona Audcent-Ross$^{3,4}$, Holger Baumgardt$^2$, Kenji Bekki$^{3,4}$\\ \
1: Research School of Astronomy and Astrophysics, The Australian National University, Cotter Road, Weston Creek, ACT 2611, Australia; \\
2: School of Mathematics and Physics, University of Queensland, Qld, 4072, Australia;\\
3: School of Physics, University of Western Australia, 35 Stirling Highway, Crawley, WA, 6009, Australia; \\
4: International Centre for Radio Astronomy Research, ICRAR M468, 35 Stirling Highway, Crawley, WA, 6009, Australia; \\
5: Centre for Astrophysics \& Supercomputing, Swinburne University of Technology, Mail number H30, PO Box 218, Hawthorn, Victoria 3122, Australia\\
* sarah@sarahsweet.com.au}
\date{Released 2015 Xxxxx XX}

\pagerange{\pageref{firstpage}--\pageref{lastpage}} \pubyear{2014}

\date{\today}

 \maketitle
\begin{abstract} 
We present DEIMOS multi-object spectroscopy (MOS) of 22 star-forming dwarf galaxies located in four gas-rich groups, including six newly-discovered dwarfs. Two of the galaxies are strong tidal dwarf galaxy (TDG) candidates based on our luminosity-metallicity relation definition. We model the rotation curves of these galaxies. Our sample shows low mass-to-light ratios (M/L=0.73$\pm0.39M_\odot/L_\odot$) as expected for young, star-forming dwarfs. One of the galaxies in our sample has an apparently strongly-falling rotation curve, reaching zero rotational velocity outside the turnover radius of $r_{turn}=1.2r_e$. This may be 1) a polar ring galaxy, with a tilted bar within a face-on disk; 2) a kinematic warp. These scenarios are indistinguishable with our current data due to limitations of slit alignment inherent to MOS-mode observations. We consider whether TDGs can be detected based on their tidal radius, beyond which tidal stripping removes kinematic tracers such as H$\alpha$ emission. When the tidal radius is less than about twice the turnover radius, the expected falling rotation curve cannot be reliably measured. This is problematic for as much as half of our sample, and indeed more generally, galaxies in groups like these. Further to this, the H$\alpha$ light that remains must be sufficiently bright to be detected; this is only the case for three (14\%) galaxies in our sample. We conclude that the falling rotation curves expected of tidal dwarf galaxies are intrinsically difficult to detect.\\
\end{abstract}


\section{Introduction} 

Occupying the low-mass end of the galaxy mass function, dwarf galaxies are more numerous than their giant counterparts {\citep[e.g.][]{Bell2003,Oppenheimer2010}}, and as such can be an important tracer of environmental effects {\citep[e.g.][]{Drinkwater2003}}. Moreover, the contribution by dwarf galaxies to hierarchical assembly and accretion onto giant galaxies is a vital part of our { understanding of galaxy formation \citep[e.g.][and many observational and theoretical works since]{Searle1978}.} In $\Lambda$CDM cosmology, galaxies form in haloes of cold dark matter {\citep[DM,][]{Peebles1965,Press1974,Blumenthal1984}}. These galaxies assemble into larger galaxies within still-larger haloes, which contribute a high M/L ratio and flat rotation curve for all galaxy masses {\citep[e.g.][]{Navarro1998,Klypin2002}}. At the same time, continual enrichment of the inter-stellar medium means that larger galaxies have higher metallicities \citep[e.g.][]{Lequeux1979,Tremonti2004}.

Galaxy groups are ideal laboratories for the study of dwarf galaxies, as it is at this density that the environment begins to contribute to their evolution. Moreover, the harsh conditions caused by galaxy clusters are not present, so that galaxy harassment is less frequent \citep{Lewis2002,Gomez2003}. Interactions between giant galaxies in such groups can cause the necessary Jeans instabilities for small galaxies to form in the tidal debris \citep{Bournaud2010}. These so-called tidal dwarf galaxies (TDGs) share the high metallicity of the giant galaxies from which they formed, so are unusually metal-rich for their small size \citep[e.g.][]{Mirabel1992,Duc2000,Weilbacher2003}. {They also have no or little non-baryonic DM, though may contain a significant fraction of baryonic DM \citep{Bournaud2007}. Notwithstanding, they have a low M/L ratio \citep[e.g.][]{Braine2001} and are expected to have} a falling, mass-follows-light rotation curve \citep[e.g.][]{Bournaud2010,Duc2014}. 
Understanding the fraction of dwarf galaxies that form in a tidal manner instead of in the traditional hierarchical assembly paradigm is crucial to instructing cosmological simulations.

Most TDGs discovered to date have been detected due to their location in the tidal streams in which they form \citep[e.g.][]{Mirabel1992,Duc2000,Weilbacher2003,Duc2011,Duc2014}. However, the same streams that identify these objects as TDGs also indicate tidal distortion of their velocity fields, so that dynamical masses cannot be reliably measured, nor can the presence or absence of DM be determined \citep{Casas2012}. For example, dwarf galaxies in obviously interacting systems such as the Perseus Cluster have comparable M/L ratios ($\sim$120 M$_\odot$/L$_\odot$) to the CDM dwarf satellites of the Milky Way \citep{Penny2009}.
The exception to this problem is where tidal streams are old and faded, as in \citet{Duc2014}, and the TDG has had time to reach dynamical equilibrium. 

{ The aim of this paper is firstly to detect TDGs based on their elevated metallicity and falling rotation curve alone, without the presence of tidal streams that confuse kinematic measurements.}
To that end, here we present rotation curves and metallicities for a sample of star-forming dwarfs in gas-rich galaxy groups, many of which do not have obvious evidence of tidal streams. Section 2 outlines our sample selection, observations and data processing. In Section 3 we present our results, including rotation curve modelling, mass-to-light ratio, and one rapidly-falling rotation curve. 
{ Finding that the rotation curves of many galaxies in the sample are distorted even without the presence of obvious tidal streams, in Section 4 we discuss the survival and detectability of TDGs in the group environment; this is the second main aim of the paper}. Section 5 concludes the paper. The Appendices contain notes on and rotation curves for individual galaxies in the sample.

\section{Sample Selection, Observations and Data Processing}

Our sample consists of dwarf galaxies within the star-forming, gas-rich groups known as Choirs \citep{Sweet2013}. These fifteen groups of four or more H$\alpha$ emission line galaxies were detected in the Survey for Ionisation in Neutral Gas Galaxies \citep[SINGG][]{Meurer2006}, a follow-up H$\alpha$ imaging survey to the HI Parkes All Sky Survey \citep[HIPASS][]{Barnes2001}.

{ In \citet{Sweet2014} we presented ANU 2.3-m WiFeS \citep{Dopita2007} integral field spectroscopy for 53 galaxies in eight of these Choir groups. We measured spatially-integrated metallicities for that sample in order to identify a sample of tidal dwarf galaxy (TDG) candidates based on elevated metallicity for their $R$-band magnitude. To confirm whether or not they are true TDGs we set out to measure their rotation curves; a falling rotation curve would demonstrate the absence of a dark matter halo as expected for such galaxies. However, with the small telescope aperture we do not enjoy sufficient signal-to-noise to allow resolved kinematic analyses of the faint dwarfs in the sample.}

We therefore obtained Keck DEIMOS \citep{Faber2003} spectroscopy for four Choir groups on 2013 February 11{ ; two groups (J1051-17 and J1403-06) from our WiFeS sample and two new groups (J0443-05 and J1059-09), being sources with observable right ascensions during the allocated observing date.}
We used the 1200L grating with central wavelength 5950 \AA\/ and the GG400 order-blocking filter. Each group was observed for six 1200-s exposures.
Mask and slit placements are shown in Appendix C. We chose the mask position angle to facilitate observing as many of our Choir member galaxies as possible. 
The remaining spare mask area was { used to allocate slits to other sources suggestive of some net H$\alpha$ emission in the SINGG imaging but not already identified, with the intention of detecting new group member galaxies.} 
This selection expanded our sample to  $\sim$100 objects per mask.

Data processing was conducted using the DEIMOS DEEP2 data reduction pipeline\footnote{http://www2.keck.hawaii.edu/inst/deimos/pipeline.html}. This pipeline is optimised for compact sources observed with a central wavelength of around 7700 \AA, so we employed Evan Kirby's modification\footnote{http://www2.keck.hawaii.edu/inst/deimos/calib\_blue.html} to enable processing of our bluer wavelengths. We also modified the sky subtraction routines to give more flexibility for extended sources, as most of our galaxies are sufficiently extended to fill their slits. In these cases we selected a section of another slit that contains some sky to perform a non-local sky subtraction. While there is some residual sky emission due to variations in the observed sky spectrum caused by slit angle and width differences and path through the optics and detector, the residual sky emission does not overlap with the H$\alpha$ for the redshifts of our groups, so the result is adequate for velocity measurements. 

We used our own software to extract a portion of the spectrum at the expected H$\alpha$ wavelength for each row of binned pixels, and fit a single Gaussian profile to the peak to measure redshift. There is no evidence for multiple components at the resolution of our data. The peak location error from the $\chi^2$-minimisation fit is used to derive the 1-$\sigma$ { observed} velocity errors quoted herein. The H$\alpha$ width and flux and continuum flux were also measured for each bin. In order to measure the true profile width, the spectrum cutout was deconvolved with a line spread function (LSF) measured from a nearby bright sky line. Importantly, we noticed that the LSF varied significantly with the tilt and location of the slit on the mask, so we chose a sky line in each slit for the LSF measurement. Heliocentric velocity corrections were calculated using the {\sc iraf} package {\sc rvcorrect}. 
Our measurements confirm that six of the potential member galaxies in two groups lie at the group redshift; these are included in Table~\ref{members}. { We used the methods outlined in \citet{Meurer2006} to measure these new galaxies' extinction-corrected R-band magnitudes, surface brightness profiles and effective radii from the SINGG imaging.}

We observed one spectrophotometric standard star for each mask and performed flux calibration in the standard manner.

We measured the integrated metallicity of each galaxy by collapsing spectra in the spatial direction and fitting a Gaussian profile to these integrated spectra to measure strong emission line fluxes. 
The wavelength range of DEIMOS 1200L grating with our central wavelength covers most of the necessary strong emission lines for using the \citet{Dopita2013} [N{\sc ii}]/[S{\sc ii}] vs. [O{\sc iii}]/[S{\sc ii}] diagnostic. This is the same calibration as used in \citet{Sweet2014}; we give a discussion of our reasons for choosing it in that paper. Briefly, this diagnostic gives consistent results with recombination and electron temperature methods; the necessary lines are available in most of our sample; and is not degenerate, clearly separating ionization parameter and metallicity{: the [N{\sc ii}]/[S{\sc ii}] ratio is sensitive to metallicity, while the [O{\sc iii}]/[S{\sc ii}] ratio tells the dependence on ionization parameter.}
The metallicities derived from the DEIMOS data are not corrected for reddening in most cases, because most do not have measurable H$\beta$, with the line either lost in the noise of the blue end, or fallen off of the chip. In any case, the diagnostic is not very dependent on reddening: because the [N{\sc ii}] and [S{\sc ii}] lines are nearby in wavelength, the [N{\sc ii}]/[S{\sc ii}] ratio will not vary much with reddening.

For the galaxies with all three strong emission lines available, we use the methods described in \citet{Sweet2014} to interpolate for metallicity and ionization parameter log($q$). { For the galaxies with only [N{\sc ii}] and [S{\sc ii}] available, we} roughly estimated the ionization parameter log($q$) based on a polynomial fit to log($q$) as a function of { $R$-band absolute magnitude} M$_R$ of the DEIMOS galaxies for which all three strong lines are available; log($q$) = 10.836 + 0.3593M$_R$ + 8.361x10$^{-3}$xM$_R^2$. We then used [N{\sc ii}]/[S{\sc ii}]  and log(q) to estimate 12 + log(O/H) by inspection of the same diagnostic. These measurements are therefore less reliable and have a nominal 0.5 dex error bar to show this. While the  [N{\sc ii}]/[S{\sc ii}] ratio alone has a similar scatter to the [N{\sc ii}]/H$\alpha$ calibration \citep[e.g.][]{Marino2013}, the method just described has the added benefit of providing log($q$), which gives a better constraint on the metallicity.
 The metallicity measurements are catalogued in Table~\ref{members}.

\section{Results and Analysis}

{ Our first step towards identifying TDGs aside from their location in a tidal stream is to select candidates based on their elevated metallicity; the second is to confirm the absence of dark matter by measuring a falling rotation curve. In this section we employ that method, firstly presenting our sample of dwarf galaxies on the luminosity-metallicity relation, and then modelling their rotation curves. We then measure mass-to-light ratios for those galaxies where this can be reliably obtained. The section finishes with a discussion of one galaxy with an apparently strongly-falling rotation curve.}

\subsection{Luminosity-metallicity relation}
\begin{figure}
\centerline{
\includegraphics[width=1\linewidth]{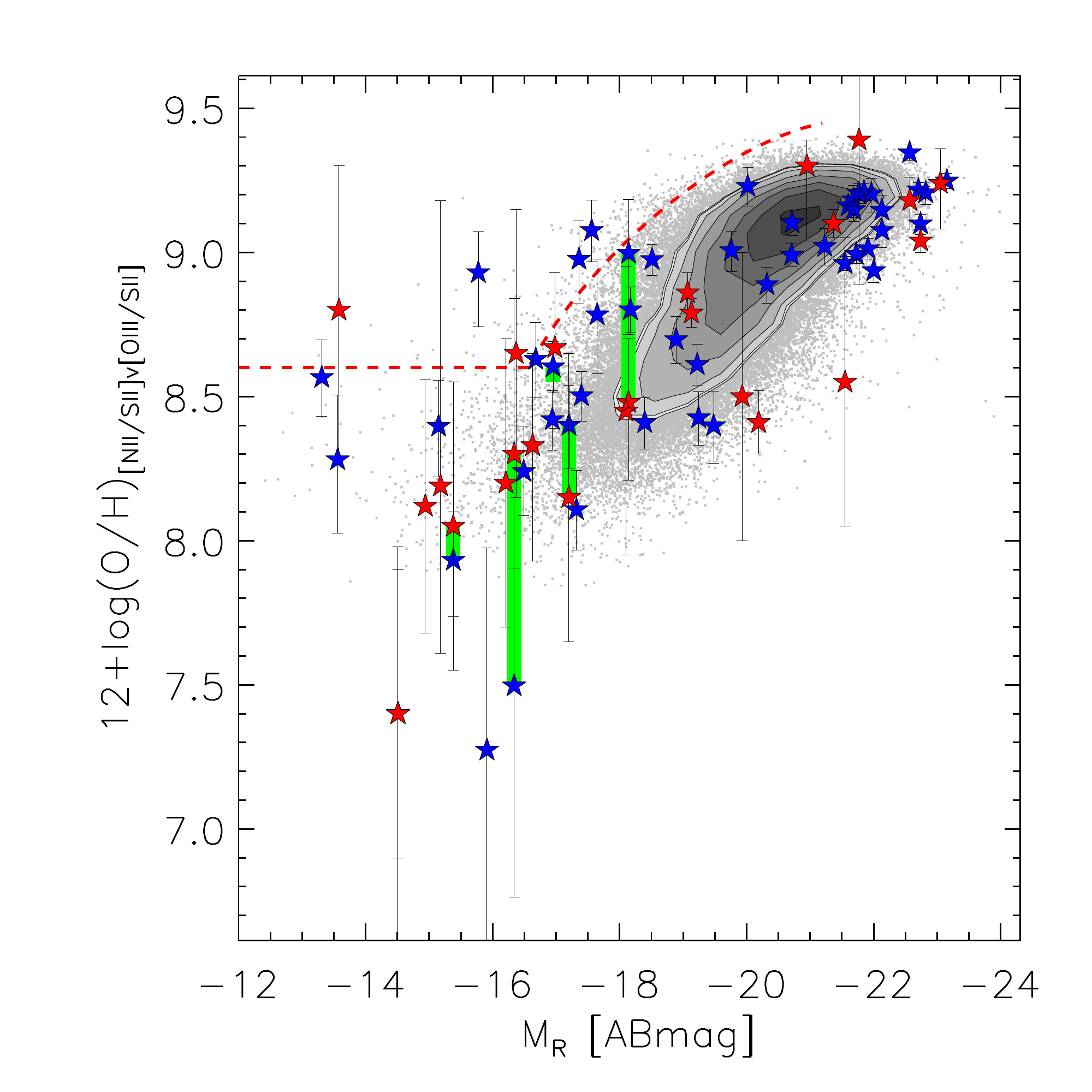}
}
\caption{Luminosity-metallicity relation for the full sample of galaxies. Blue stars are existing WiFeS measurements from \citet{Sweet2014}. Red stars are DEIMOS measurements. Green bars connect WiFeS and DEIMOS metallicities for five of the six galaxies that have measurements from both instruments; both measurements for J1403-06:S4 at $M_R=-14.51$ are equal. 
 Grey points and contours depict SDSS star-forming galaxies, while the red, dashed line indicates our strong TDG candidate diagnostic. Galaxies above this line are more than 3 sigma above the typical metallicity for their luminosity and hence candidate TDGs.
\label{LZ}}
\end{figure}

In Figure~\ref{LZ} we plot the luminosity-metallicity relation for our full sample of WiFeS \citep{Sweet2014} and DEIMOS (this work) measurements. In general, the galaxies with DEIMOS measurements are consistent with the portion of our sample for which we have WiFeS measurements, and with the SDSS relation defined in \citet{Sweet2014}. In that paper we identified a metallicity floor for low luminosity galaxies in SDSS. There is a suggestion of the metallicity floor being continued to fainter magnitudes by the new measurements.

We select TDG candidates in the DEIMOS data based on our definition from \citet{Sweet2014} (12+log(O/H)$>$8.6 and more than 3$\sigma$ above SDSS); these are J1051-17:g11 and J1403-06:g1. However, J1403-06:g1 lies in the halo of its host spiral, and is probably experiencing strong tidal interactions, so is not detectable as a TDG based on its velocity profile (see Section~\ref{detected}). 

Six of the galaxies in our sample have both WiFeS and DEIMOS metallicity measurements. In the J1403-06 group these metallicities agree to within quoted errors, even without [O{\sc iii}] for the DEIMOS measurements. However, the J1051-17 group members where we can compare the data generally do not have consistent metallicities between DEIMOS and WiFeS, separated by up to 2$\sigma$. This is partly due to the missing [O{\sc iii}] line for some of the DEIMOS measurements, but more generally because bright sky lines fall on the [S{\sc ii}] lines at the distance of this group. There is some unavoidable sky residual for both instruments due to the non-local sky subtraction employed. The sky residuals therefore contribute to the [S{\sc ii}] flux measurements and skew the metallicities for this group. We expect that the DEIMOS sky subtraction discussed above may be worse than for the WiFeS observations, where we used a nod-and-shuffle technique which minimised systematics related to optics and detector position. 

\subsection{Rotation curve modelling}

\begin{figure*}
\centerline{
\includegraphics[width=1\linewidth]{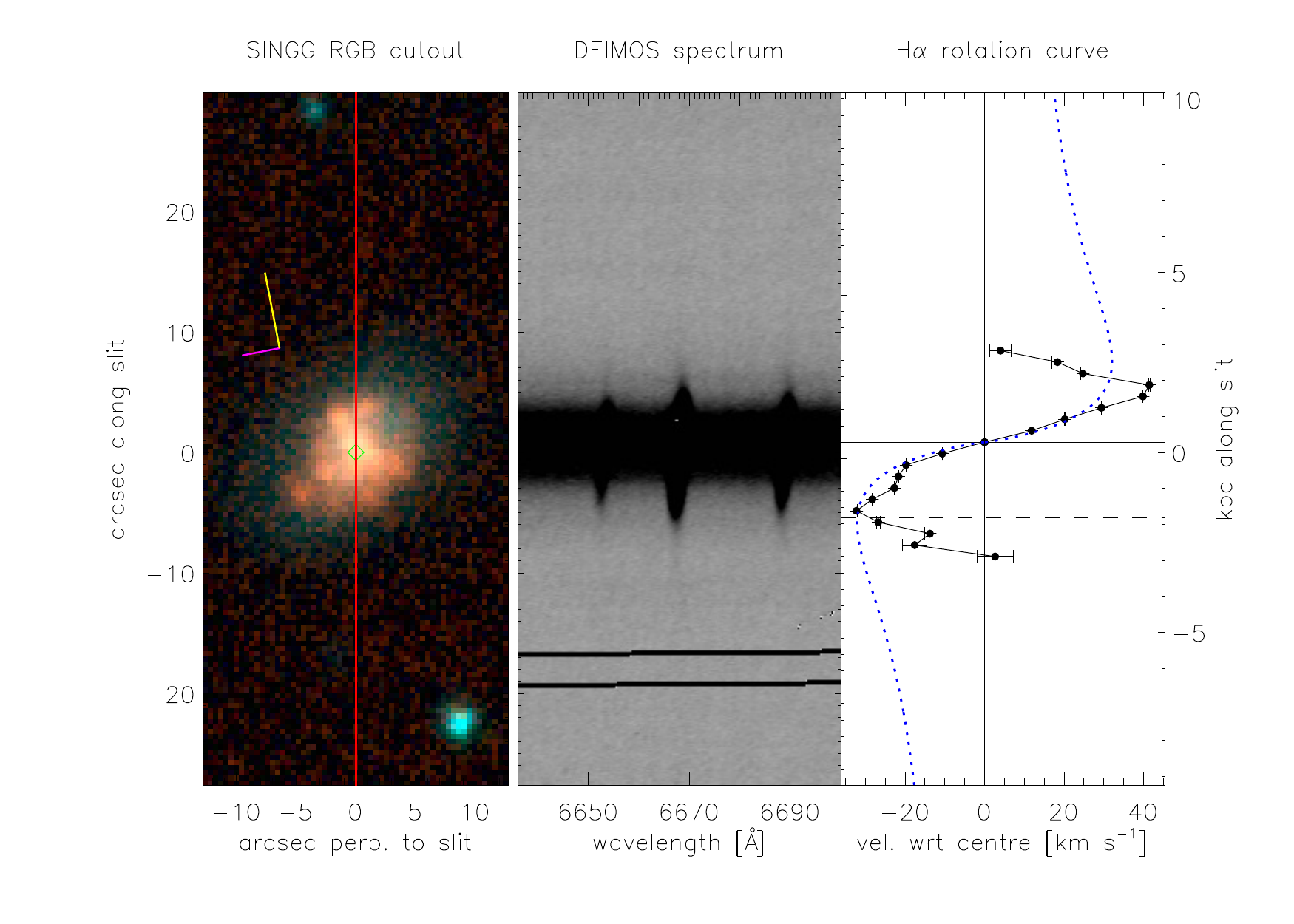}
}
\caption{SINGG image, DEIMOS spectrum and rotation curve for HIPASS J0443-05:S4. (a) The left-most panel is the SINGG RGB image (R = R-band, G = narrow-band filter, B = net H$\alpha$). Stars appear cyan and HII regions appear red in this colour scheme. The red, vertical lines indicate the position of the slit, which runs the length of this panel. The compass shows N in yellow and E in magenta. The green diamond shows the optical (SINGG) centre of the object, corresponding to (0,0) on this plot. 
(b) The central panel is a cutout from the DEIMOS spectrum, showing H$\alpha$ and [N{\sc ii}] emission lines. This covers the same angular height as the left-hand panel.
(c) The right-hand panel is also the same angular height as the other two panels, with a matching kiloparsec scale on the right-hand axis. It contains the observed (uncorrected) rotation curve in black dots. The crosshairs are centered on the continuum flux peak centre, which is not necessarily the same as the optical centre in the left-hand panel. The dashed lines show the expected turnover radius 2$r_0$ = 1.2$r_e$, assuming an exponential disk.
The blue, dotted curve is a mass-follows-light rotational velocity prediction based on SINGG photometry and fitted to the observed rotation curve within the turnover radius. See text for details.
\label{0443_S4}}
\end{figure*}

We constructed rotation curves along the DEIMOS slit for each of the galaxies in our sample. We defined the zero velocity point as the velocity at the bin with the maximum continuum flux. This centre is a reasonable match to optical SINGG isophotal measurement of the centre for most of the galaxies in the sample. For J0443-05:S3 there are bad rows very near the centre of the galaxy, so for this galaxy we fit the centre by eye, assuming a symmetrical mass distribution. 
 The rotation curve for one of the the more interesting galaxies in our sample, J0443-05:S4, is shown in Figure~\ref{0443_S4} and discussed further in Section \ref{special}. Rotation curves for the other galaxies in our sample, along with notes on each, are presented in Appendices A and B.

We test for the absence of DM by modelling the rotation curves of these galaxies. If the observed rotation curve is consistent with a mass-follows-light profile instead of being flat, then DM is not required to explain the rotational velocity. 
We assume an exponential disk and use the analytical prescriptions of \citet{Freeman1970} for this modelling: the predicted rotational velocity for a mass-follows-light system is given by $v_{rot}^2 = \pi G \mu_0 r^2(I_0K_0 - I_1K_1)$. Here, $\mu_0 = M/L \times S_0$ is the central surface density, $M/L$ is the mass-to-light ratio and $S_0$ is the central surface luminosity. $I_{0(1)}$ and $K_{0(1)}$ are modified Bessel functions of the zeroth (first) kind evaluated at $\frac{r}{2r_0}$, where $r_0$ is the scale length. This mass-follows-light profile is akin to a Keplerian falloff beyond the turnover radius $r_{turn}  = 2r_0 = 1.2r_e$, where $r_e$ is the $R$-band effective radius. (This is for a purely exponential galaxy; for a bulge-dominated galaxy the turnover will be at a smaller multiple of $r_e$.) Our measurements of $\mu_0$ and $r_0$ are derived assuming pure exponential disks from the effective surface brightness and radius measurements in the $R$-band, as described in \citet{Meurer2006}.

We corrected for inclination using the optical axial ratio $b/a$ from SINGG photometry, where $b$ is the semi-minor axis and $a$ is the semi-major axis, and the angle of inclination $i$ is given by $\rm{cos}(i-3\degree) = \sqrt{((b/a)^2 - 0.2^2)/(1-0.2^2)}$, the factor of 0.2 is a correction for disk thickness \citep{Tully1977} and 3\degree\/ is an empirical correction for the difference in the flattening of stellar vs. HI disks \citep{Aaronson1980}. We also corrected for angle of misalignment $\phi$ between slit position angle and SINGG optical major axis position angle. The misalignment is a function of the chosen slit mask position angle and the constraints of the instrument; the slit position angle (PA) must be $5\degree<| PA|<30\degree$ with respect to the mask, limiting the choice of position angle. The model rotation curve obtained from fitting the optical data is { divided} by the total correction of sin($i$)cos($\phi$). 

{ We do not have sufficient information to model the asymmetric drift. While we have measured the H$\alpha$ velocity dispersion at each point in our spectra, this emission is strongly affected by HII regions. It is not clear that these are in virial equilibrium, since their dynamics are thought to be dominated by short-lived ($\lesssim$ 10 Myr) expansion from ionisation, stellar winds and supernovae \citep{Shopbell1998,Clarke2002}. The nature of the diffuse ionised gas H$\alpha$ between HII regions is also unclear, and in many cases is a signature of outflows such as galactic winds \citep{Oey2007,Rodriguez2008}. We also do not have sufficient information about the neutral and molecular ISM (which is being ionised to become H$\alpha$) to determine its contribution: our HI data has insufficient resolution, and we have no molecular observations. We can estimate the order of magnitude of the { asymmetric drift} as follows (noting that this estimate is an upper limit, since the dispersion in the H$\alpha$ emission is overestimated for the reasons just mentioned). The asymmetric drift correction term $\sigma_D$ as a function of radius $r$ is given by 
$\sigma_D^2 = -r\sigma^2
[\partial{ln(\Sigma_g)}/\partial{r} + 2\partial{ln(\sigma)}/\partial{r} - \partial{ln(h_z)}/\partial{r}]$, 
where $\sigma$ is the Gaussian sigma of the 1D H$\alpha$ velocity profile measured as described in the previous section, $h_z$ is the vertical scale height of the disk and $\Sigma_g$ is the gas surface density. For most of the galaxies in the sample there is no obvious dependence of $\sigma$ on the radius, so we adopt a single median value for each galaxy, with a sample mean of 20 km s$^{-1}$.
We also assume that $h_z$ is constant with radius, so the correction term becomes simply $\sigma_D^2 = -r\sigma^2[\partial{ln(\Sigma_g)}/\partial{r}]$. 
Making a further assumption that the ionised gas distribution is similar to the neutral gas and star distributions allows fitting an exponential profile to the natural log-scaled continuum flux as a function of $r$. Doing so results in a mean asymmetric drift correction of around 6 km s$^{-1}$ at maximum velocity. As this is a simplified estimation only, we do not include this correction in our further analysis.
}

\subsection{Rotation curve quality}
Reliable mass-to-light ratios are dependent on good quality observations. Unfortunately, not many of the rotation curves in this sample are simple to analyse: several are disturbed, having asymmetric velocity profiles, and others have multiple SF regions or possible counter-rotating cores. In our case we restrict our M/L analysis to the galaxies that meet the following criteria:
\begin{enumerate}
\item sin($i$)cos($\phi$) $>$ 0.4, so there is neither a large correction for galaxy inclination nor misalignment between the slit and galaxy position angle; and,
\item sufficiently large and bright so that the photometry is reliable (this cutoff is effectively between galaxies in the original SINGG sample and fainter galaxies identified in this work); and,
\item not strongly disturbed in appearance (e.g. Fig.~\ref{1403_S4}).
\end{enumerate}
The resulting sample, which we label Sample A, consists of ten dwarf galaxies, out of 22 for which we have DEIMOS measurements. The galaxies within Sample A are indicated in Table~\ref{members}.

The fraction of our sample that meets criterion (i) is 60\% (all of the galaxies in Sample A, plus three that fail criteria (ii) and (iii)). We can calculate the expected fraction that meets the first criterion as follows. The position angle misalignment is given by $\phi = PA-30\degree$, where the galaxy position angle with respect to the mask is $30\degree\leq PA\leq 90\degree$. and $\phi = 0\degree$ for $0\degree\leq PA<30\degree$. For a randomly-oriented galaxy, a PA of 45$\degree$ then gives a median $\phi$ = 15$\degree$. Half of a sample of such galaxies should therefore have $\phi \leq 15\degree$.
 Solving sin($i$)cos(15$\degree$) $>$ 0.4 for $i$ gives a minimum inclination of $23\degree$, below which the amplitude in the rotation curve rapidly becomes too low to measure. For a randomly-selected sample, the frequency of any given {cos(inclination)} should be constant, so up to { $\sim$cos(23$\degree$) = 92\%} of galaxies are more edge-on than this. The resulting expected fraction of galaxies that meets our first criterion is therefore { 0.5 x 0.92 = 46\%}; considerably lower than the 60\% in our sample. { The position angle calculations rely on our assumption of disk-like systems. However, the irregular galaxies may be better modelled as triaxial systems, which are more prolate. This means that there is a lower chance of observing these to be circular and measuring low, face-on inclinations in the disk-like model \citep{vandenBergh1988}, so much of the difference between the expected and observed fraction of edge-on galaxies is} likely attributable to triaxiality of the irregular galaxies. { In addition to this, we also expect that there are} group effects causing a non-random orientation in position angles in our sample. In particular, there is a hint of a stream of dwarf galaxies in J1051-17 (see Fig.~\ref{1051}), which is to be discussed in a forthcoming paper (Kilborn et al., 2015, in prep.).

\subsection{M/L ratios}

We test for the presence of DM by calculating M/L ratios through fitting model curves to observed rotational velocities as described above. A high M/L ratio implies the presence of DM and a normal hierarchical formation mechanism. In general the M/L ratios in Sample A are low, but not dramatically so, at a mean M/L ratio of 0.73$M_\odot/L_\odot$ { and standard error on the mean of $\pm0.39M_\odot/L_\odot$}.
Low M/L ratios in general simply confirm that these are young galaxies forming stars. Some of our wider sample does not have sufficient signal at radii $> r_{turn}$, so that our modelling calculates lower limits on masses and mass-to-light ratios. However, this does not significantly affect Sample A, which all have reliable measurements within $r_{turn}$ and are well-modelled.
 
\begin{figure}
\centerline{
\includegraphics[width=1\linewidth]{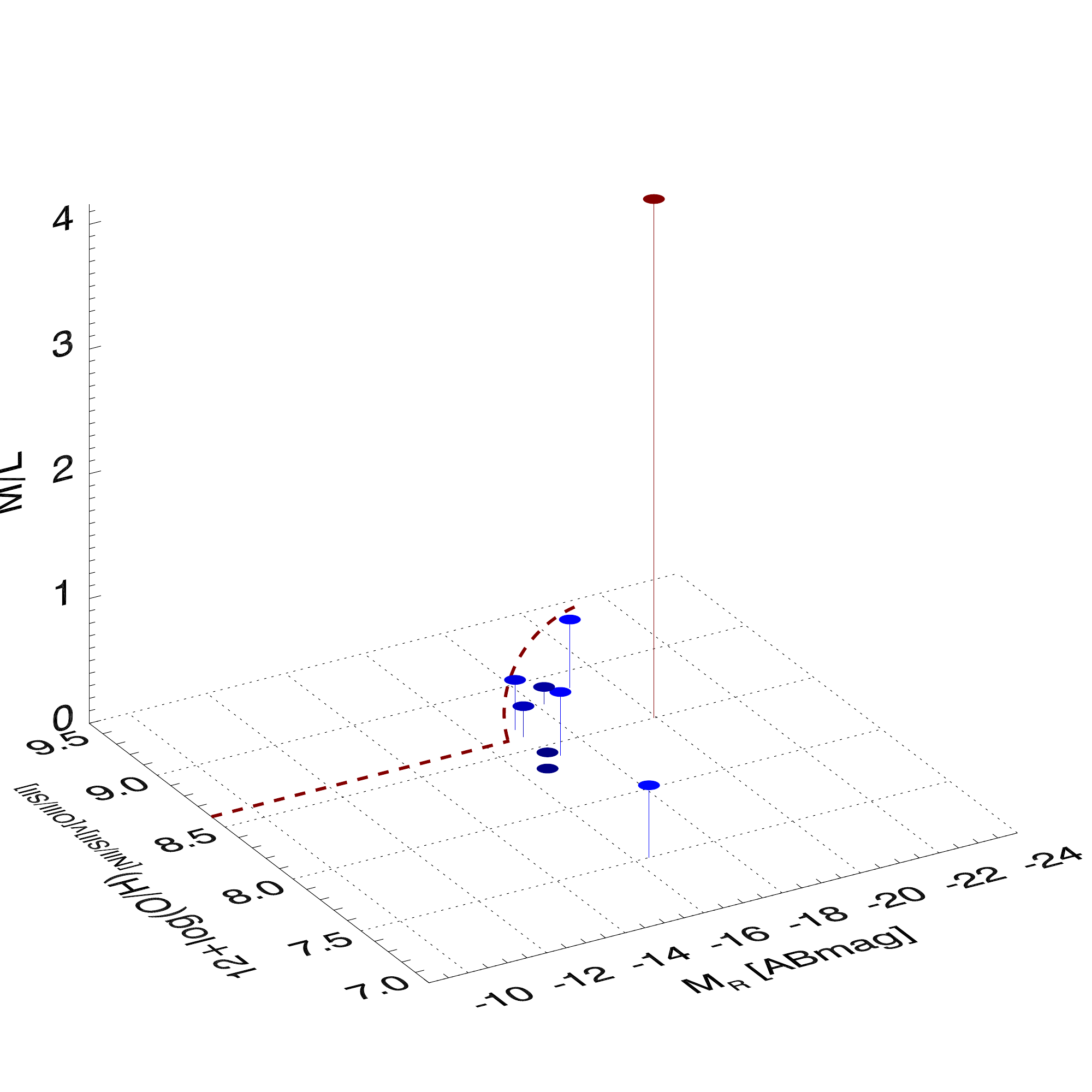}
}
\caption{Mass-to-light ratio as a function of the luminosity-metallicity relation for the galaxies in our sample that have reliable mass-to-light ratios. The data points are colour-coded by mass-to-light ratio, with red being the highest and blue the lowest in this sample. The red, dashed line is the strong TDG candidate diagnostic line as defined in \citet{Sweet2014}.
\label{LZM}}
\end{figure}

 We plot M/L ratios as a function of the luminosity-metallicity relation using Sample A in Fig.~\ref{LZM}. While sparse, the data suggest a trend towards higher mass-to-light ratios with higher luminosity. This is consistent with the view that the luminosity-metallicity relation arises from the deeper potential well of larger galaxies, which makes them more able to retain metal-rich supernovae ejecta \citep{Gibson1997,Kauffmann2003}. 
 
Unfortunately, neither of the strong TDG candidates identified above is part of Sample A, so we cannot include either of them in this analysis. In essence this is because neither has a well-behaved mass-follows-light rotation curve: J1051-17:g11 has no detectable H$\alpha$ on the southern semi-major axis, and a large correction for inclination; J1403-06:g1 has a disturbed rotation curve.

An improved strategy for identifying TDGs in this manner is: 1) measure metallicities with low-resolution spectra of as many galaxies as possible in a field, preferring those with easiest (edge-on) inclinations; and 2) measure rotational velocity with medium-resolution spectra of a subset of these, either optimising slit orientation for the high-metallicity candidates, or using an integral field spectrograph. In this way the effects of large inclination / position angle corrections can be minimised. However, this strategy does not take into account tidally disturbed or stripped matter at the outskirts of target galaxies, leading us to consider that problem in Section~\ref{detected}.

\subsection{\label{special}The apparently strongly-falling rotation curve of J0443-05:S4}

{ As discussed in the previous sections, no galaxy in our sample that is a \emph{strong} TDG candidate, based on elevated metallicity, exhibits a falling rotation curve. Relaxing the metallicity criterion somewhat, our next} best candidate for a TDG based on metallicity and rotation curve is HIPASS J0443-05:S4 shown in Fig.~\ref{0443_S4}. In terms of metallicity, this galaxy is 12+log(O/H)=8.86 and $R$-band magnitude M$_R$=-19.07, placing it within the metallicity-luminosity relation defined by the SDSS control sample. However, it is near a giant spiral S2 which has a metallicity just 0.32 dex higher at 12+log(O/H)=9.18 and M$_R$=-22.57. We note that these differences in magnitude and metallicity are similar to those measured in the M31-M32 system \citep[$\Delta$mag = 4.43; $\Delta$metallicity = 0.22 dex;][]{Richer1998}, for which a tidal \emph{encounter} has been proposed \citep[e.g.][]{Faber1973,Bekki2001b,Choi2002}. 

The measured rotation curve is falling rapidly beyond the turnover radius $r_{turn}$, { indicating that dark matter is not \emph{required}} to explain its observed velocity ({ though it is not disallowed}). In fact, the rotation curve shows a significant down-turn \emph{below} the mass-follows-light profile shown in blue, even reaching zero at the outskirts of the galaxy. This galaxy is near our nominal detection limit of $r_{tidal} = 2r_{turn}$, indicating that there could be some tidal warping of the rotation curve. However, the symmetry of this system indicates it is not subject to extreme tidal forces, so we do not believe that tides are causing the severe down-turn. The rapidly-falling rotation curve is not consistent with the predictions for a TDG, and the lack of evidence for tides corroborates this. The error bars (derived from the peak location error in the $\chi^2$-minimisation fit to the H$\alpha$ line) indicate that low S/N is not responsible for the unusual shape of this rotation curve. Nor can the downturn be attributed to instrumental signatures or residual sky lines; the same shape is observed for other emission lines in this system, which fall in other locations on the detector.

{ Clearly, a simplified disk model is not adequate to describe this galaxy. We propose two scenarios that may explain} the stronger than Keplerian fall-off exhibited by the measured rotation curve: 1) polar ring galaxy, with a tilted inner structure and face-on outer disk; 2) kinematic twist due to a warped disk. 

Firstly, we consider a polar ring galaxy, where the central, rising { observed} velocity profile belongs to a tilted inner bar, and the falling profile to a face-on outer disk with no measurable rotation. The observed morphology hints at a central bar, though more detailed imaging is required to confirm this. A polar ring structure like this is unusual in a dwarf \citep[cf.][]{deRijcke2013}, but could conceivably arise as a tilt to the galaxy's existing disk, triggered by an interaction with the neigbouring giant galaxy S2, if not the typical (for giant polar ring galaxies) method of accreting material from the nearby galaxy \citep{Athanassoula1985}. The size of the bar relative to the disk in this scenario is reminiscent of the morphological (cf. dynamical) bar of the LMC\footnote{To be clear, the LMC is not considered by most to be a TDG.}, as is the dwarf-giant separation ($\sim50$ kpc in both cases). \citet{vanderMarel2014} recently measured the LMC (stellar) rotation curve, but due to the high degree of scatter in the stellar velocities there is no clear trend in the outskirts of the galaxy with which to compare J0443-05:S4.
		
Our second hypothesis is that we are observing a kinematic twist or warp, where the kinematic position angle varies smoothly with radius \citep{Krajnovic2008}. Kinematic and isophotal warps are caused by bars, often occurring with spiral arms \citep[e.g. VCC0523, ][]{Rys2013} or shells \citep[e.g. Figure 5 of ][]{Emsellem2006}. For a good visual example of an isophotal warp see \citet[][]{Elmegreen1996}, especially their Figure 1 (NGC1300 I). A warp of this magnitude would cause the line of nodes (region of zero velocity with respect to the systemic velocity) to lie at the edges of a poorly-aligned slit, mimicking a falling rotation curve such as we observe for J0443-05:S4. Indeed, the velocity profile along the minor axis of NGC1068 exhibits a falling rotation curve similar to ours \citep{Emsellem2006}. 

To distinguish between these two degenerate scenarios we require additional data, in the form of integral field spectroscopy.

\section{Discussion: Can the falling rotation curves of TDGs be detected?}\label{detected}

In order to dynamically confirm a tidal dwarf galaxy by the absence of DM, when the M/L ratio is not known \emph{a priori}, it is necessary to measure its velocity to sufficient radii to detect any fall in velocity, consistent with the predicted mass-follows-light model curve. In principle, because of the gradual, Keplerian fall-off, this translates to measurements of at least twice the turnover radius $r_{turn}=1.2r_e$ for a pure exponential disk. This measurement becomes difficult at an exponential rate if it is assumed that our kinematics tracer of H$\alpha$ emission has the same profile as continuum light \citep{Dopita1994,Ryder1994}.
In practice, however, there are two bigger issues with measuring the rotation curve of a TDG. 

Firstly, the H$\alpha$ emission may be clumpy and more centrally-concentrated than the continuum light, making the measurement more difficult than predicted. We define the dynamical confirmation radius $r_{det}$ { as the maximum radius at which sensible fits to the emission in the spectra can be obtained (this translates to a S/N ratio of $\sim$ 3). This radius is} a function mostly of the distribution of the H$\alpha$ light but also influenced by the proximity to other SF regions (e.g. see the discussion for J1051-17:S7 in Appendix A). Identification becomes an issue if $2r_{turn} \gtrsim r_{det}$. This can be visualised in Figure~\ref{0443_S4} and Figures~\ref{0443_S3} to~\ref{1403_g1}. If the H$\alpha$ light is \emph{less} centrally-concentrated than the continuum light, then this criterion works in our favour because the measurement at large radii is easier to make than for a simple exponential. Either way, one must assume that the velocity of HII regions is representative of the rotation curve. This will often be the case, even if the HII region is very large and non-central, as seen in \citet{Richards2014}.

Secondly, the more critical issue is that the strong tidal fields that form a TDG, together with the low interior mass density expected for a DM-free TDG torn from the low-density outskirts of galaxies, lead to the tidal stripping of the outer baryonic matter of the TDG. With all other things equal, a normal galaxy with the same stellar mass as a TDG should be better able to survive tidal stretching because of its dark matter which increases its cohesion. Tidal stripping makes the measurement impossible when $2r_{turn} \gtrsim r_{tidal}$, where $r_{tidal} = 0.4d(m/M)^{1/3}$ is the tidal radius of the dwarf galaxy \citep[for a fluid, triaxial satellite, as in][]{Shu1982}; $d$ = distance of closest approach, $m$ = mass of dwarf galaxy, $M$ = mass of giant host galaxy.  
It is worth noting that the velocity field will be disturbed even before the proximity criterion for stripping is reached, for instance \citet{Bekki2011} showed that the rotation curve of a MW-type galaxy changes over time with tidal heating. As discussed in \citet{Renaud2009}, tidal fields are in fact compressive as well as destructive; while the destructive stretching acts to pull galaxies apart, it is the compression mechanism that is responsible for the formation of TDGs. \citet{Aguilar1986} also pointed out that while strong tidal encounters decrease the effective radius of a galaxy, weak encounters puff up the galaxy instead, e.g. by galaxy harassment \citep{Moore1996}. This disturbance is also problematic for a sound measurement of a rotation curve. However, we do not attempt to quantify this here for the sake of simplicity in this analysis.

We calculate $r_{tidal}$ for each galaxy in our sample, assuming $d = \sqrt{2}d_{proj}$, where $d_{proj}$ = projected distance to the nearest large galaxy, and the factor of $\sqrt{2}$ is based on a 45\degree\/ projection angle. This assumes that the deprojected distance is the distance of closest approach. It is clearly very likely that this is not in fact the closest approach because the dwarf galaxies are more likely either on their way towards or away from the nearby giant, so will be closer in the future or have already been in the past. Therefore we are overestimating the tidal radius.
For this calculation we also assume a M/L ratio for `dwarfs' (galaxies with $M_R > -22$) of 1 M$_\odot$/L$_\odot$, and 4 M$_\odot$/L$_\odot$ for `giant' ($M_R < -22$) galaxies, consistent with \citet{McGaugh1998,Toloba2011,Kalinova2013}. This gives a mass ratio $m/M$ of $1/4(l/L)$, where $l$ and $L$ are the luminosity of the dwarf and giant galaxy respectively, so that $r_{tidal} = 0.4\sqrt{2}d_{proj}(l/4L)^{1/3}$.

\begin{figure}
\centerline{
\includegraphics[width=1\linewidth]{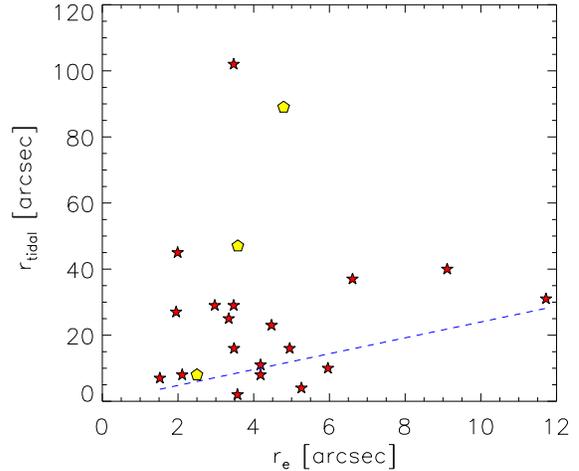}
}
\caption{Tidal radius vs. effective radius for our sample of galaxies. See text for calculation of tidal radius. The dashed line indicates nominal detection limit of 2$r_{turn}$ = 2.4$r_e$. Below this limit it is not possible to dynamically confirm a falling rotation curve because baryons are tidally stripped down to 2$r_{turn}$ (the radius at which a falling rotation curve may be detected). Yellow pentagons represent the galaxies in our sample that have measurable H$\alpha$ beyond 2$r_{turn}$; the galaxies that are not measurable at that radius are shown as red stars.
\label{rturn}}
\end{figure}

In Fig.~\ref{rturn} we plot $r_{tidal}$ vs. $r_{e}$; the dashed line indicates the required radius of $2r_{turn}$. Four of the 22 galaxies in our sample fall below this line, indicating that these could not be dynamically confirmed as tidal dwarf galaxies because no baryons should remain bound where the rotation curve is falling. A further five to six galaxies are borderline. Others in this sample could also be unidentifiable, given that the tidal radius is likely overestimated. Clearly, a large $r_{tidal}$ is required for a galaxy to be dynamically confirmed to be a TDG; that is, the dwarf must be far from the nearest giant. However, the further away from the giant, the less likely the dwarf is to be formed in a tidal manner (c.f. J0443-05:S3, which has the largest $r_{tidal}$ in the sample, and a flat (DM-rich) rotation curve, so is not a TDG). Further to this, we reiterate that the $2r_{turn}$ detection limit only indicates where the falling-velocity baryons do not remain bound. If a dwarf galaxy lies above this line then those baryons remain bound, but there must also be sufficiently bright H$\alpha$ to meet the detectability radius $r_{det}$ criterion before the galaxy can be detected as a TDG. Even with these DEIMOS observations only three (14\%) galaxies have $r_{det} \gtrsim 2r_{turn}$ (shown as yellow pentagons in Fig.~\ref{rturn}): J0443-05:S3, J1051-17:S5, J1059-09:S7. None of these has a falling rotation curve, so none is a TDG.

This effect is problematic for TDG candidates in the literature that have been identified based on their location within tidal streams, because the strong tides distort the rotation curves and dynamical M/L ratios. This is clear in the work by \citet{MendesdeOliveira2001}, who used velocity gradients to ascertain whether or not star-forming regions within tidal tails in Stephan's Quintet would remain bound, forming TDGs. None of the rotation curves shows a fall-off in velocity, and all are severely disturbed by the tidal field. The M/L ratios measured are inflated (5-73 M$_\odot$/L$_\odot$), requiring DM in opposition to expectations for TDGs. It may be argued that one can do better with HI measurements, probing to larger radii than H$\alpha$. Declining rotation curves are seen in some galaxies observed in HI including DDO 154 (\citet{Carignan1998,Hoffman2001}; however cf. \citet{deBlok2008}) and NGC 300 \citep{Westmeier2011}. In the case of DDO154, this measurement was possible because the galaxy is isolated and hence is not tidally truncated. However, this mass includes a substantial DM component and therefore rules out a tidal origin for this galaxy.
Measurements of falling rotation curves remain difficult, except for a few select cases.

\section{Conclusions}
In this paper we presented DEIMOS observations of a sample of 22 star-forming dwarf galaxies in gas-rich groups. After prioritising our known group members in the slit mask design, we placed spare slits on as many sources as possible, with preference to galaxy-like photometry. In doing so we identified six additional small galaxies across two of the groups. 

We measured the metallicity of those galaxies within our sample that have the necessary strong emission lines and found two new very strong TDG candidates (J1051-17:g11 and J1403-06:g1). 

We constructed rotation curves for the dwarf galaxies in our sample and modelled them with a mass-follows-light fit to the central regions of each galaxy. All but one of the galaxies show signs of rotation with a mean of 39.3 $\rm{km~s^{-1}}$ at $r_{turn}$, but most of the velocity profiles are disturbed so that a mass-follows-light profile does not fit the data, and neither does a mass-follows-light plus a DM component. The generally disturbed nature of the velocity profiles indicate that these galaxies are tidally perturbed.
 
M/L ratios in our sample are low (0.73$\pm0.39M_\odot/L_\odot$), indicating that the stellar populations in these galaxies are young, consistent with their high rates of star-formation. There is some suggestion of a trend of M/L ratio with luminosity in our sample, with fainter galaxies having lower mass-to-light ratios.
 
One galaxy in our sample, J0443-05:S4, has an apparently strongly-falling rotation curve, reaching zero velocity at the outskirts of the galaxy. We propose that we may be observing either 1) a polar ring galaxy, with a tilted inner structure and face-on outer disk; 2) a kinematic twist due to a warped disk, with the line of nodes falling within the edges of the slit.

Even with very high sensitivity DEIMOS data, it remains difficult to convincingly measure the falling rotation curve of a TDG, due to both physical and observational effects. Observationally, the limitations of slit - galaxy position angle alignment severely constrain our ability to reliably measure kinematics of all galaxies in a group. To overcome these observational effects, integral field unit spectroscopy should be employed. For DEIMOS with a constraint of $\pm 30\degree$, we have reliable measurements for 60\% of our sample. This is considerably greater than the predicted 37.5\%, suggesting that the position angles of the galaxies are aligned, perhaps due to group effects. 

Physically, many of the rotation curves in our sample are disturbed due to recent interaction, or are not smooth due to having multiple star-forming regions. In addition to this, the outskirts of many of the dwarfs may be tidally stripped by interactions with neighbouring galaxies. As much as half of our sample could be affected by this. Even in the absence of tidal stripping, H$\alpha$ light rapidly becomes progressively fainter beyond the turnover radius. Only 14\% of our sample has detectable H$\alpha$ light at sufficient radii to measure any fall in rotation curve; none of these has a falling rotation curve, so none is a TDG. It seems that falling rotation curves expected of TDGs can be detected only rarely, if at all.

\section*{Acknowledgements}
{We thank Peter Weilbacher and the anonymous referee for reviewing this paper.}

We acknowledge funding support from the UQ-UWA Bilateral Research Collaboration Award and the Australian Research Council Discovery Project.

Some of the data presented herein were obtained at the W.M. Keck Observatory, which is operated as a scientific partnership among the California Institute of Technology, the University of California and the National Aeronautics and Space Administration. The Observatory was made possible by the generous financial support of the W.M. Keck Foundation. The authors wish to recognise and acknowledge the very significant cultural role and reverence that the summit of Mauna Kea has always had within the indigenous Hawaiian community.  We are most fortunate to have the opportunity to conduct observations from this mountain.

This research has made use of the NASA's Astrophysics Data System.

This research has made use of the NASA/IPAC Extragalactic Database (NED) which is operated by the Jet Propulsion Laboratory, California Institute of Technology, under contract with the National Aeronautics and Space Administration.

This research has made use of the USNO Image and Catalogue Archive
   operated by the United States Naval Observatory, Flagstaff Station
   (http://www.nofs.navy.mil/data/fchpix/).

\clearpage
\onecolumn
\begin{centering}
\begin{landscape}
\begin{longtable}{|l|l|l|l|l|l|l|l|l|l|l|l|l|l|l|l|l|l|l|l|l|l|l|}
\caption{Measured quantities for group member galaxies, including new group members. Columns (1) to (11).
\label{members}}\\
\hline 
\multicolumn{1}{|c|}{(1)}&\multicolumn{1}{|c|}{(2)}&\multicolumn{1}{|c|}{(3)}&\multicolumn{1}{|c|}{(4)}&\multicolumn{1}{|c|}{(5)}&\multicolumn{1}{|c|}{(6)}&\multicolumn{1}{|c|}{(7)}&\multicolumn{1}{|c|}{(8)}&\multicolumn{1}{|c|}{(9)}&\multicolumn{1}{|c|}{(10)}&\multicolumn{1}{|c|}{(11)}\\
\multicolumn{1}{|c|}{HIPASS+}&\multicolumn{1}{|c|}{RA}&\multicolumn{1}{|c|}{Dec}&\multicolumn{1}{|c|}{reff}&\multicolumn{1}{|c|}{a/b}&\multicolumn{1}{|c|}{PA}&\multicolumn{1}{|c|}{corr.}&\multicolumn{1}{|c|}{M$_R$}&\multicolumn{1}{|c|}{$r_0$}&\multicolumn{1}{|c|}{$I_0$}&\multicolumn{1}{|c|}{V$_{hel}$}\\
\multicolumn{1}{|c|}{}&\multicolumn{1}{|c|}{[h m s]}&\multicolumn{1}{|c|}{[d m s]}&\multicolumn{1}{|c|}{['']}&\multicolumn{1}{|c|}{}&\multicolumn{1}{|c|}{[$\degree$]}&\multicolumn{1}{|c|}{}&\multicolumn{1}{|c|}{[mag]}&\multicolumn{1}{|c|}{[kpc]}&\multicolumn{1}{|c|}{[L$_\odot$/kpc]}&\multicolumn{1}{|c|}{[km/s]}\\
\hline
\hline
\endfirsthead
\caption[]{Measured quantities for group member galaxies, including new group members. Columns (1) to (11).
}\\
\hline 
\multicolumn{1}{|c|}{(1)}&\multicolumn{1}{|c|}{(2)}&\multicolumn{1}{|c|}{(3)}&\multicolumn{1}{|c|}{(4)}&\multicolumn{1}{|c|}{(5)}&\multicolumn{1}{|c|}{(6)}&\multicolumn{1}{|c|}{(7)}&\multicolumn{1}{|c|}{(8)}&\multicolumn{1}{|c|}{(9)}&\multicolumn{1}{|c|}{(10)}&\multicolumn{1}{|c|}{(11)}\\
\multicolumn{1}{|c|}{HIPASS+}&\multicolumn{1}{|c|}{RA}&\multicolumn{1}{|c|}{Dec}&\multicolumn{1}{|c|}{reff}&\multicolumn{1}{|c|}{a/b}&\multicolumn{1}{|c|}{PA}&\multicolumn{1}{|c|}{corr.}&\multicolumn{1}{|c|}{M$_R$}&\multicolumn{1}{|c|}{$r_0$}&\multicolumn{1}{|c|}{$I_0$}&\multicolumn{1}{|c|}{V$_{hel}$}\\
\multicolumn{1}{|c|}{}&\multicolumn{1}{|c|}{[h m s]}&\multicolumn{1}{|c|}{[d m s]}&\multicolumn{1}{|c|}{['']}&\multicolumn{1}{|c|}{}&\multicolumn{1}{|c|}{[$\degree$]}&\multicolumn{1}{|c|}{}&\multicolumn{1}{|c|}{[mag]}&\multicolumn{1}{|c|}{[kpc]}&\multicolumn{1}{|c|}{[L$_\odot$/kpc]}&\multicolumn{1}{|c|}{[km/s]}\\
\hline
\hline
\endhead
\hline
\multicolumn{11}{r}{Continued on next page}
\endfoot
\endlastfoot
J0443-05:S3          & 04 44 11.67 & -05 14 38.31 & 04.79$\pm$0.19 &1.86&19&0.883&-19.93$\pm$0.19&1.809&2.3E+8&4591\\
J0443-05:S4          & 04 44 05.54 & -05 25 46.50 & 04.95$\pm$0.12 &1.6&120&0.537&-19.07$\pm$0.12&1.048&5.9E+8&4774\\
J1051-17:S3          & 10 51 35.94 & -16 59 16.80 & 06.61$\pm$0.05 &1.02&74&0.048&-18.14$\pm$0.05&2.232&4.7E+7&5969\\
J1051-17:S4& 10 51 26.01 & -17 05 03.61 & 03.48$\pm$0.09 &1.4&164&0.661&-16.34$\pm$0.09&0.837&7.5E+7&5465\\
J1051-17:S5          & 10 51 50.91 & -16 58 31.64 & 03.58$\pm$0.06 &1.75&29&0.865&-17.20$\pm$0.06&0.842&1.9E+8&5465\\
J1051-17:S6          & 10 51 42.78 & -17 06 34.59 & 02.11$\pm$0.04 &1.29&40&0.422&-16.95$\pm$0.04&0.492&3.0E+8&5648\\
J1051-17:S7          & 10 51 33.36 & -17 08 36.63 & 04.18$\pm$0.12 &1.53&49&0.802&-16.94$\pm$0.12&0.799&1.2E+8&5374\\
J1051-17:S8   & 10 51 25.92 & -17 08 16.44 & 04.47$\pm$0.10 &3.07&63&0.927&-18.17$\pm$0.04&0.804&7.5E+8&5294\\
J1051-17:g04&10 51 39.679&-17 03 34.16&02.97$\pm$0.14&1.06&43&0.21&-18.10$\pm$0.05&1.45&5.0E+7&5535\\
J1051-17:g07&10 51 43.698&-17 01 42.99&03.34$\pm$0.09&1.12&42&0.3&-16.21$\pm$0.08&0.95&4.0E+7&6166\\
J1051-17:g11&10 51 40.051&-16 57 30.94&01.95$\pm$0.13&1&47&0.027&-16.37$\pm$0.11&0.28&1.3E+9&5371\\
J1051-17:g13&10 51 41.602&-17 05 20.16&01.52$\pm$0.16&1.05&42&0.21&-14.94$\pm$0.21&0.685&8.0E+6&5577\\
J1051-17:g15&10 51 33.286&-17 08 19.17&05.26$\pm$0.85&1.05&42&0.21&-15.18$\pm$0.81&2.63&4.0E+6&5225\\
J1059-09:S2          & 10 59 06.77 & -09 45 04.38 & 11.72$\pm$0.24 &1.36&131&0.056&-20.19$\pm$0.24&5.271&4.7E+7&8013\\
J1059-09:S5          & 10 59 30.98 & -09 44 25.26 & 09.11$\pm$0.13 &2.84&75&0.968&-19.94$\pm$0.13&2.973&2.4E+8&7926\\
J1059-09:S7          & 10 59 21.31 & -09 47 50.49 & 02.50$\pm$0.15 &1.59&115&0.167&-19.13$\pm$0.15&0.761&1.2E+9&7862\\
J1059-09:S8          & 10 59 01.73 & -09 52 46.76 & 03.47$\pm$0.40 &1.81&155&0.702&-16.98$\pm$0.40&1.028&9.4E+7&8260\\
J1059-09:S9  & 10 58 44.69 & -09 53 28.60 &01.99$\pm$0.08&1.39&45&0.604&-16.63$\pm$0.08&0.63&2.0E+8&8475\\
J1059-09:S10   & 10 59 02.64 & -09 53 19.90&03.47$\pm$0.01&1.2&40&0.746&-20.95$\pm$0.01&4.11&5.0E+7&8219\\
J1403-06:S3& 14 03 13.48 & -06 06 24.17 & 04.18$\pm$0.85 &1.03&14&0.158&-15.38$\pm$0.85&0.448&7.5E+7&2753\\
J1403-06:S4& 14 03 34.62 & -06 07 59.27 & 05.96$\pm$0.86 &1.43&123&0.731&-14.51$\pm$0.86&0.961&1.4E+7&2671\\
J1403-06:g1&14 03 22.475&-06 00 44.24&03.57$\pm$0.36&2.04&46&0.76&-13.58$\pm$1.39&2.51&6.3E+6&2692\\
\hline
\end{longtable}
Columns: (1) SINGG name; (2) right ascension; (3) declination; 
(4) effective radius; (5) ratio of major to minor axes; (6) position angle; (7) velocity correction sin($i$)cos($\phi$); (8) $R$-band extinction-corrected magnitude; (9) scale length; (10) central surface luminosity; (11) heliocentric velocity.
\clearpage
\addtocounter{table}{-1}
\begin{longtable}{|l|l|l|l|l|l|l|l|l|l|l|l|l|l|l|l|l|l|l|l|l|l|l|}
\caption[Columns (12) to (22).]{Measured quantities for group member galaxies, including new group members. Columns (12) to (22).
}\\
\hline 
\multicolumn{1}{|c|}{(1)}&\multicolumn{1}{|c|}{(12)}&\multicolumn{1}{|c|}{(13)}&\multicolumn{1}{|c|}{(14)}&\multicolumn{1}{|c|}{(15)}&\multicolumn{1}{|c|}{(16)}&\multicolumn{1}{|c|}{(17)}&\multicolumn{1}{|c|}{(18)}&\multicolumn{1}{|c|}{(19)}&\multicolumn{1}{|c|}{(20)}&\multicolumn{1}{|c|}{(21)}&\multicolumn{1}{|c|}{(22)}\\
\multicolumn{1}{|c|}{HIPASS+}&\multicolumn{1}{|c|}{H$\beta$}&\multicolumn{1}{|c|}{[O {\sc iii}] }&\multicolumn{1}{|c|}{H$\alpha$}&\multicolumn{1}{|c|}{[N {\sc ii}] }&\multicolumn{1}{|c|}{[S {\sc ii}]}&\multicolumn{1}{|c|}{log(q)}&\multicolumn{1}{|c|}{12+log(O/H)}&\multicolumn{1}{|c|}{A}&\multicolumn{1}{|c|}{M/L}&\multicolumn{1}{|c|}{V$_{rot}$}&\multicolumn{1}{|c|}{$r_{tidal}$}\\
\multicolumn{1}{|c|}{}&\multicolumn{1}{|c|}{}&\multicolumn{1}{|c|}{}&\multicolumn{1}{|c|}{}&\multicolumn{1}{|c|}{}&\multicolumn{1}{|c|}{}&\multicolumn{1}{|c|}{}&\multicolumn{1}{|c|}{}&\multicolumn{1}{|c|}{}&\multicolumn{1}{|c|}{[M$_\odot$/L$_\odot$]}&\multicolumn{1}{|c|}{[km/s]}&\multicolumn{1}{|c|}{['']}\\
\hline
\hline
\endfirsthead
\caption[]{Measured quantities for group member galaxies, including new group members. Columns (12) to (22).
}\\
\hline 
\multicolumn{1}{|c|}{(1)}&\multicolumn{1}{|c|}{(12)}&\multicolumn{1}{|c|}{(13)}&\multicolumn{1}{|c|}{(14)}&\multicolumn{1}{|c|}{(15)}&\multicolumn{1}{|c|}{(16)}&\multicolumn{1}{|c|}{(17)}&\multicolumn{1}{|c|}{(18)}&\multicolumn{1}{|c|}{(19)}&\multicolumn{1}{|c|}{(20)}&\multicolumn{1}{|c|}{(21)}&\multicolumn{1}{|c|}{(22)}\\
\multicolumn{1}{|c|}{HIPASS+}&\multicolumn{1}{|c|}{H$\beta$}&\multicolumn{1}{|c|}{[O {\sc iii}] }&\multicolumn{1}{|c|}{H$\alpha$}&\multicolumn{1}{|c|}{[N {\sc ii}] }&\multicolumn{1}{|c|}{[S {\sc ii}]}&\multicolumn{1}{|c|}{log(q)}&\multicolumn{1}{|c|}{12+log(O/H)}&\multicolumn{1}{|c|}{A}&\multicolumn{1}{|c|}{M/L}&\multicolumn{1}{|c|}{V$_{rot}$}&\multicolumn{1}{|c|}{$r_{tidal}$}\\
\multicolumn{1}{|c|}{}&\multicolumn{1}{|c|}{}&\multicolumn{1}{|c|}{}&\multicolumn{1}{|c|}{}&\multicolumn{1}{|c|}{}&\multicolumn{1}{|c|}{}&\multicolumn{1}{|c|}{}&\multicolumn{1}{|c|}{}&\multicolumn{1}{|c|}{}&\multicolumn{1}{|c|}{[M$_\odot$/L$_\odot$]}&\multicolumn{1}{|c|}{[km/s]}&\multicolumn{1}{|c|}{['']}\\
\hline
\hline
\endhead
\hline
\multicolumn{13}{r}{Continued on next page}
\endfoot
\endlastfoot
J0443-05:S3          &&&10.5$\pm$2.1&0.8$\pm$0.4&3.4$\pm$1.5&7&8.5$\pm^{0.5}_{0.5}$&A&4.15&120&89\\
J0443-05:S4          &&28.2$\pm$8.8&174.8$\pm$1.0&46.9$\pm$0.4&69.6$\pm$1.5&6.89&8.86$\pm^{0.07}_{0.07}$&A&0.55&31.6&16\\
J1051-17:S3          &&7.2$\pm$0.7&19.0$\pm$0.2&2.8$\pm$0.2&8.6$\pm$1.0&6.8&8.48$\pm^{0.27}_{0.22}$&&&3.8&37\\
J1051-17:S4&&5.0$\pm$0.6&9.8$\pm$0.1&0.4$\pm$0.1&3.8$\pm$1.3&6.91&8.3$\pm^{0.78}_{0.54}$&A&0.58&12.2&16\\
J1051-17:S5          &&&32.4$\pm$0.4&2.9$\pm$0.3&12.4$\pm$0.7&7.13&8.15$\pm^{0.5}_{0.5}$&A&0.51&24.9&47\\
J1051-17:S6          &&&162.2$\pm$2.1&22.5$\pm$0.6&49.3$\pm$2.0&7.15&8.55$\pm^{0.5}_{0.5}$&A&0.25&8.1&8\\
J1051-17:S7          &&&&&&0&0$\pm^{0}_{0}$&A&0.03&3.8&8\\
J1051-17:S8   &&&4.7$\pm$0.1&3.9$\pm$0.2&&0&0$\pm^{0}_{0}$&A&0.14&27.4&23\\
J1051-17:g04&&&19.3$\pm$0.4&6.7$\pm$0.4&15.0$\pm$0.6&7.58&8.45$\pm^{0.5}_{0.5}$&&&14.3&29\\
J1051-17:g07&&&15.0$\pm$0.4&2.4$\pm$0.4&5.1$\pm$1.6&7.82&8.2$\pm^{0.5}_{0.5}$&&&13.8&25\\
J1051-17:g11&&&0.7$\pm$0.1&0.5$\pm$0.2&1.5$\pm$0.2&7.85&8.65$\pm^{0.5}_{0.5}$&&&20.1&27\\
J1051-17:g13&&13.0$\pm$2.3&11.9$\pm$0.2&0.7$\pm$0.2&3.4$\pm$0.6&6.99&8.12$\pm^{0.44}_{0.44}$&&&12&7\\
J1051-17:g15&&134.6$\pm$48.8&12.4$\pm$0.2&1.7$\pm$0.2&7.8$\pm$1.4&6.82&8.19$\pm^{0.58}_{0.99}$&&&0&4\\
J1059-09:S2          &&41.9$\pm$6.9&43.3$\pm$0.7&6.3$\pm$0.4&17.6$\pm$1.1&6.99&8.41$\pm^{0.11}_{0.11}$&&&33.1&31\\
J1059-09:S5          &&&&&&0&0$\pm^{0}_{0}$&A&0.28&41.4&40\\
J1059-09:S7          &10.4$\pm$0.2&13.0$\pm$2.0&65.9$\pm$1.0&14.8$\pm$0.3&23.8$\pm$0.9&6.9&8.79$\pm^{0.05}_{0.05}$&&&7.4&8\\
J1059-09:S8          &1.3$\pm$0.3&3.2$\pm$1.5&6.5$\pm$0.3&1.2$\pm$0.1&2.6$\pm$0.4&7.02&8.67$\pm^{0.28}_{0.26}$&A&0.4&14.1&29\\
J1059-09:S9  &1.4$\pm$0.1&7.5$\pm$1.4&9.0$\pm$0.3&0.7$\pm$0.1&4.0$\pm$0.4&6.91&8.33$\pm^{0.4}_{0.32}$&A&0.43&14.1&45\\
J1059-09:S10   &&4.7$\pm$0.7&&9.3$\pm$1.3&2.0$\pm$1.1&9.49&9.3$\pm^{0.26}_{0.09}$&&&0.4&102\\
J1403-06:S3&&&9.9$\pm$0.1&0.6$\pm$0.1&2.4$\pm$0.1&7.29&8.05$\pm^{0.5}_{0.5}$&&&4.7&11\\
J1403-06:S4&&&2.4$\pm$0.1&0.2$\pm$0.1&0.6$\pm$0.1&7.38&7.4$\pm^{0.5}_{0.5}$&&&4.8&10\\
J1403-06:g1&3.3$\pm$0.1&&21.5$\pm$0.2&5.7$\pm$0.2&6.9$\pm$0.1&7.56&8.8$\pm^{0.5}_{0.5}$&&&32.5&2\\
\hline
\end{longtable}
(12-16) observed flux for various emission lines in units of 10$^{-22}$ erg s$^{-1}$ cm$^{-2}$: [O {\sc iii}] $\lambda$5006.9, [N {\sc ii}] $\lambda$6583.4, [S {\sc ii}]$\lambda\lambda$6717.0+6731.3; (17) estimated ionization parameter basd on \citet{Dopita2013} interpolation if [O{\sc iii}] available, or on M$_R$-log(q) relation of other galaxies in this sample otherwise; (18) 12+log(O/H) using \citet{Dopita2013} calibration; 
(19) Membership of Sample A (based on quality of rotation curve) is indicated here by the letter `A'; (20) { $R-$band} mass-to-light ratio; (21) modelled rotational velocity at $r_{turn}$; (22) tidal radius. 
\end{landscape}
\end{centering}
\twocolumn

\clearpage
\addcontentsline{toc}{chapter}{Bibliography}
\bibliographystyle{hapj}
\bibliography{ChoirsDEIMOS}

\appendix

\section{Sample A}

\subsection{Notes on individual galaxies}

\begin{figure*}
\centerline{
\includegraphics[width=0.75\linewidth]{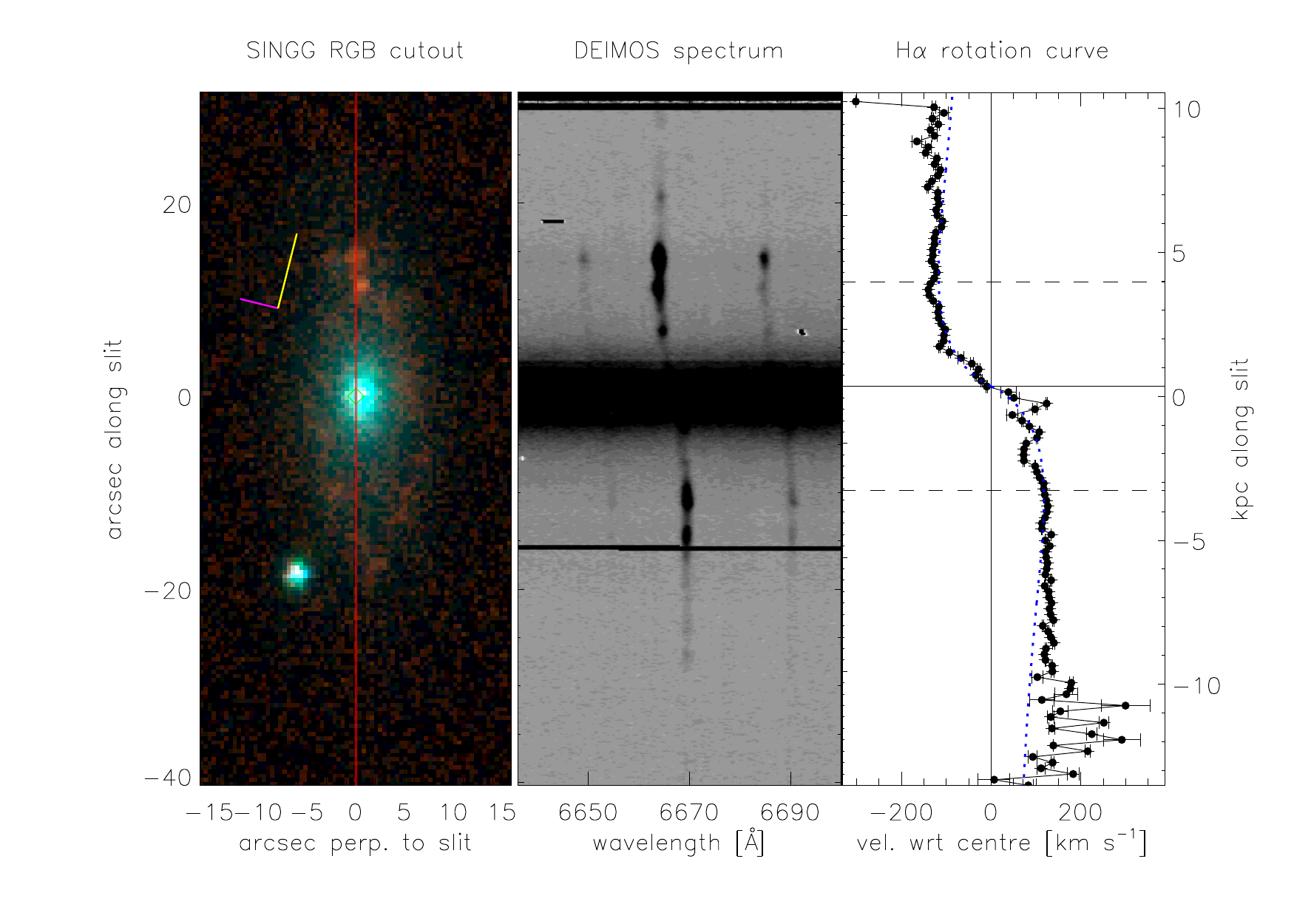}
}
\caption{J0443-05:S3. As for Figure~\ref{0443_S4}. 
\label{0443_S3}}
\end{figure*}

{ J0443-05:S3} (Figure~\ref{0443_S3}). This galaxy is the most `normal' in our sample, with a small stellar bulge (cyan) surrounded by a star-forming disk (red), and a flat rotation curve, consistent with a DM halo. Its tidal radius is much larger than its effective radius, borne out by the large radius to which we measure a typical rotation curve. It also has the highest M/L ratio in our sample at 4.15$M_\odot/L_\odot$.

{ J0443-05:S4} (Figure~\ref{0443_S4}). This galaxy has a stronger than Keplerian fall-off in its rotation curve. It is depicted in Figure~\ref{0443_S4} and is discussed in detail in Section~\ref{special}.

\begin{figure*}
\centerline{
\includegraphics[width=0.75\linewidth]{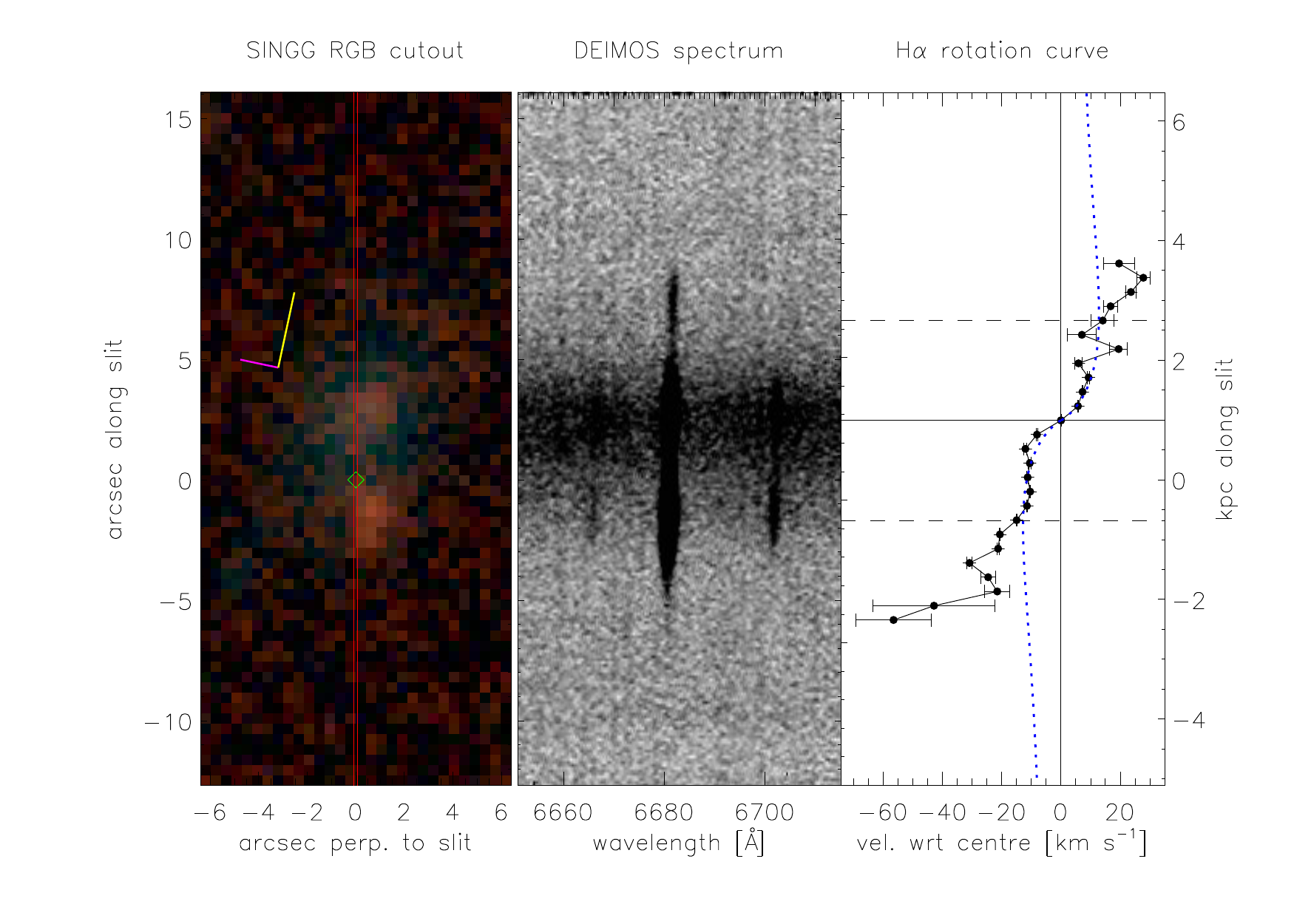}
}
\caption{J1051-17:S4. As for Figure~\ref{0443_S4}.
\label{1051_S4}}
\end{figure*}

{ J1051-17:S4} (Figure~\ref{1051_S4}). This galaxy has three separate SF regions, with velocity rising linearly with position along slit, consistent with solid body rotation. The observed rotation rises above the model predictions, suggestive of the presence of DM. It has a low M/L ratio (0.58$M_\odot/L_\odot$), consistent with most of the dwarfs in our sample.

\begin{figure*}
\centerline{
\includegraphics[width=0.75\linewidth]{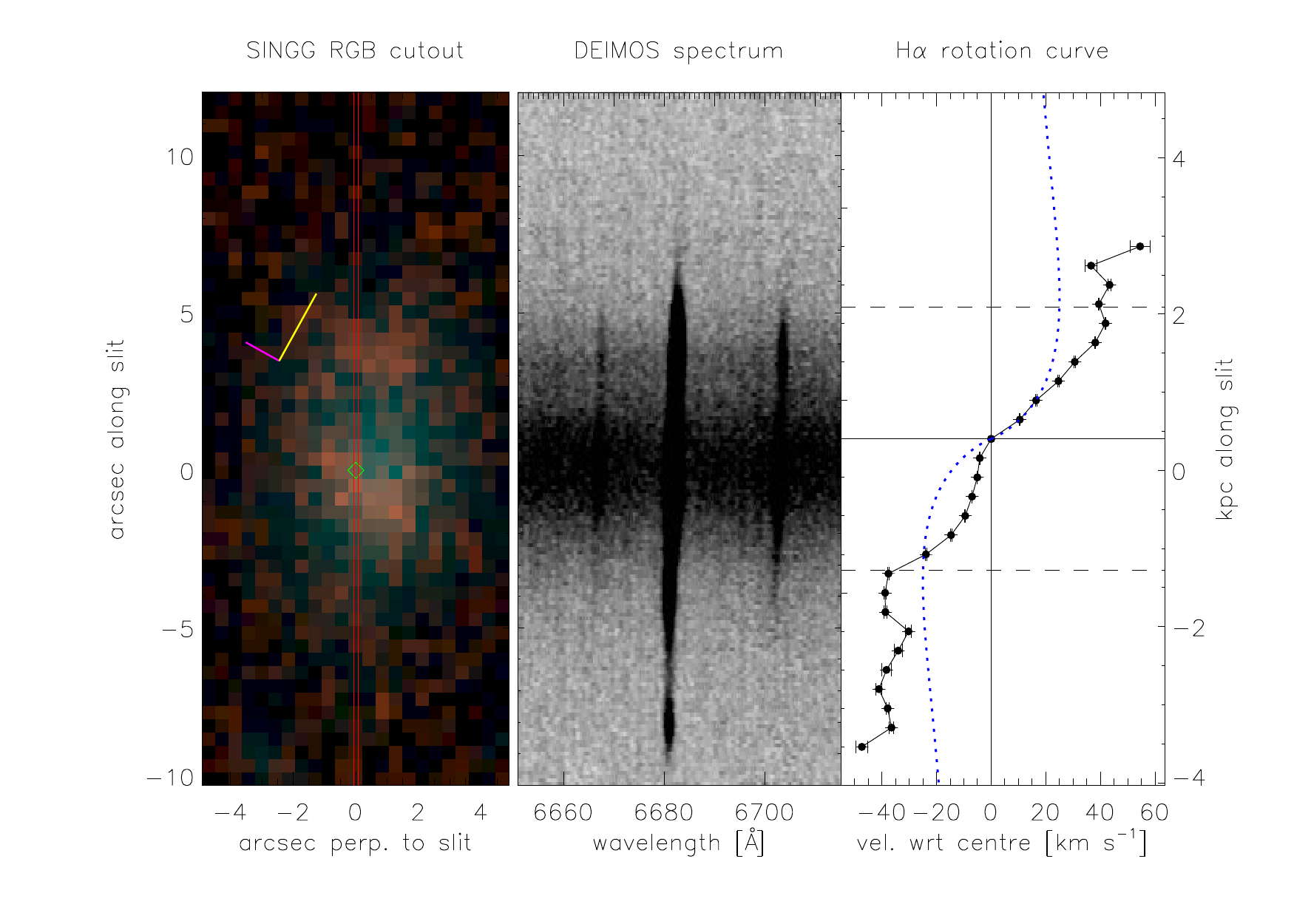}
}
\caption{J1051-17:S5. As for Figure~\ref{0443_S4}. 
\label{1051_S5}}
\end{figure*}

{ J1051-17:S5} (Figure~\ref{1051_S5}). There appear to be two SF regions in this galaxy. The velocity profile of this galaxy is not well fit by the canonical mass-follows-light rotation curve. However, this galaxy is not likely suffering from tidal effects due to its position on the $r_{tidal}-r_e$ plot.

\begin{figure*}
\centerline{
\includegraphics[width=0.75\linewidth]{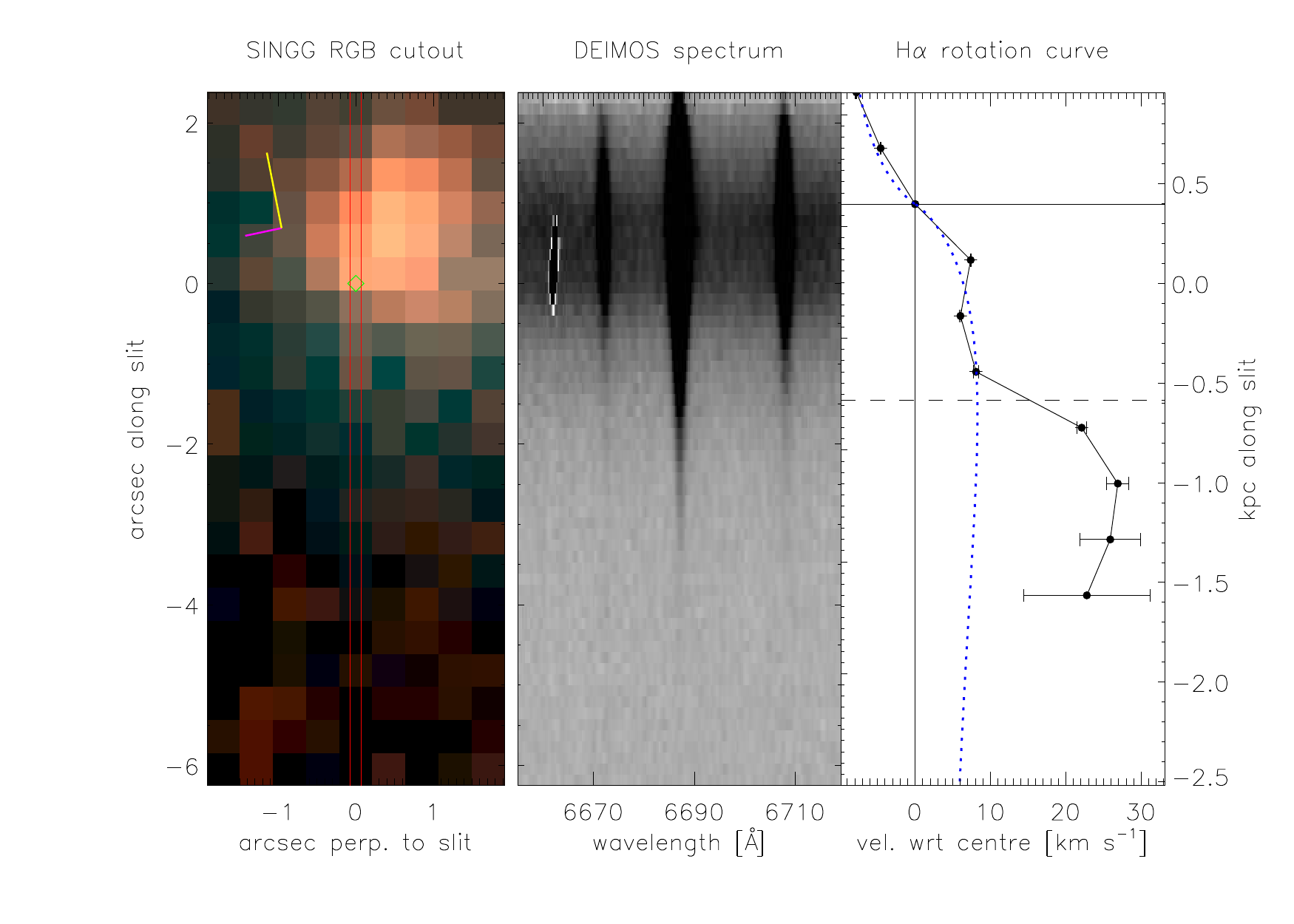}
}
\caption{J1051-17:S6. As for Figure~\ref{0443_S4}. 
\label{1051_S6}}
\end{figure*}

{ J1051-17:S6} (Figure~\ref{1051_S6}). This galaxy has a velocity similar to the disk of the nearby giant S1. The observed velocity profile of S6 is distorted beyond $r_{turn}$ by its proximity to S1.

\begin{figure*}
\centerline{
\includegraphics[width=0.75\linewidth]{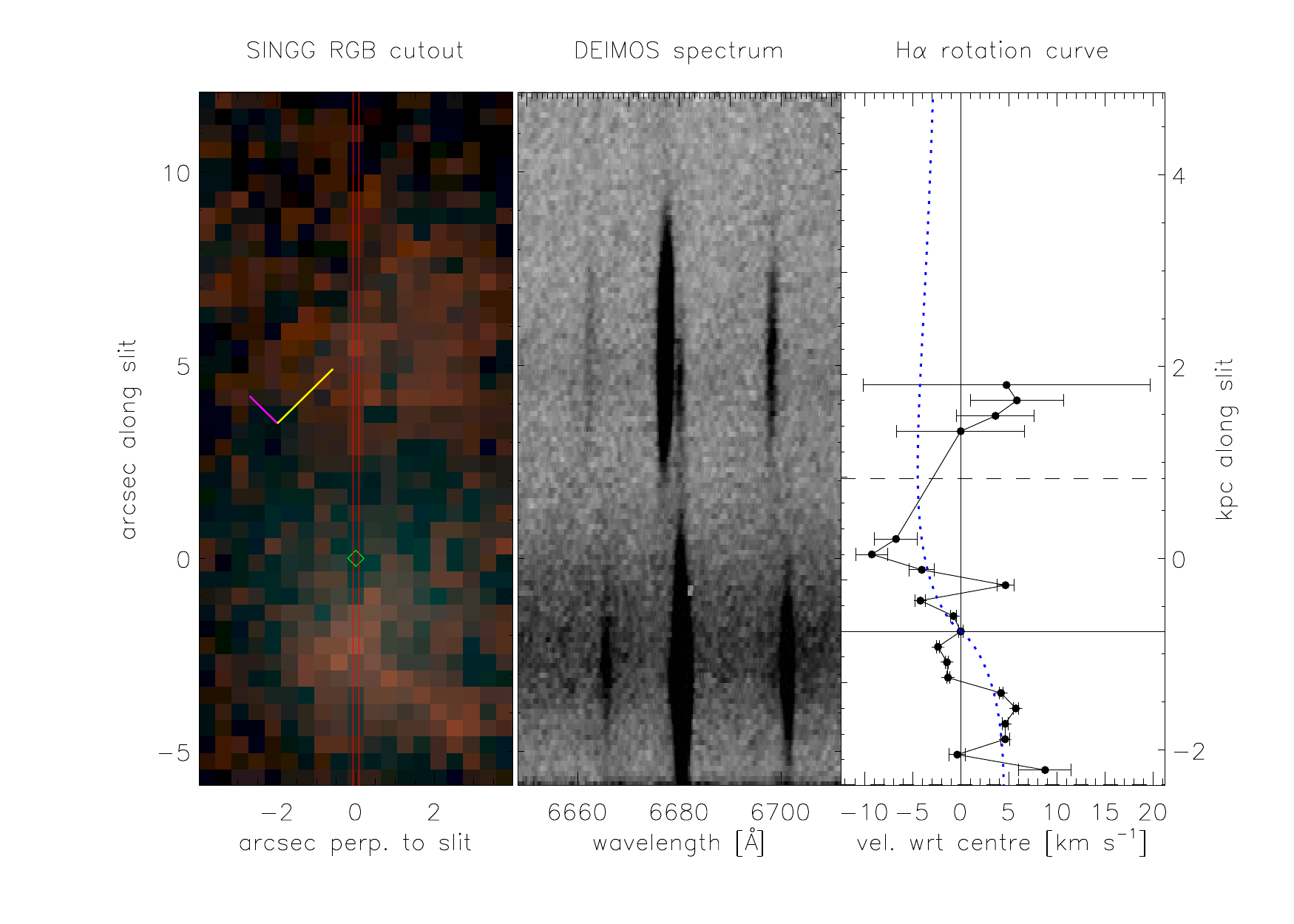}
}
\caption{J1051-17:S7. As for Figure~\ref{0443_S4}. 
\label{1051_S7}}
\end{figure*}

{ J1051-17:S7} (Figure~\ref{1051_S7}). S7 is the brighter feature to the W of the slit (that is, the bottom of this figure). Also measured in this slit is an HII region in the plane of the disk of S1 (`S7a'), towards the top of this figure. The velocity profile shown here is that of S7 only. It is disrupted, consistent with this galaxy having recently passed through the disk of S1 and possibly inducing star formation in S7a. Its velocity is clearly offset from the velocity of the neighbouring disk of S1, but the shape of the spiral arm of S1 passing near S7 is distorted and the H$\alpha$ emission is enhanced, strongly suggesting a recent interaction. 

\begin{figure*}
\centerline{
\includegraphics[width=0.75\linewidth]{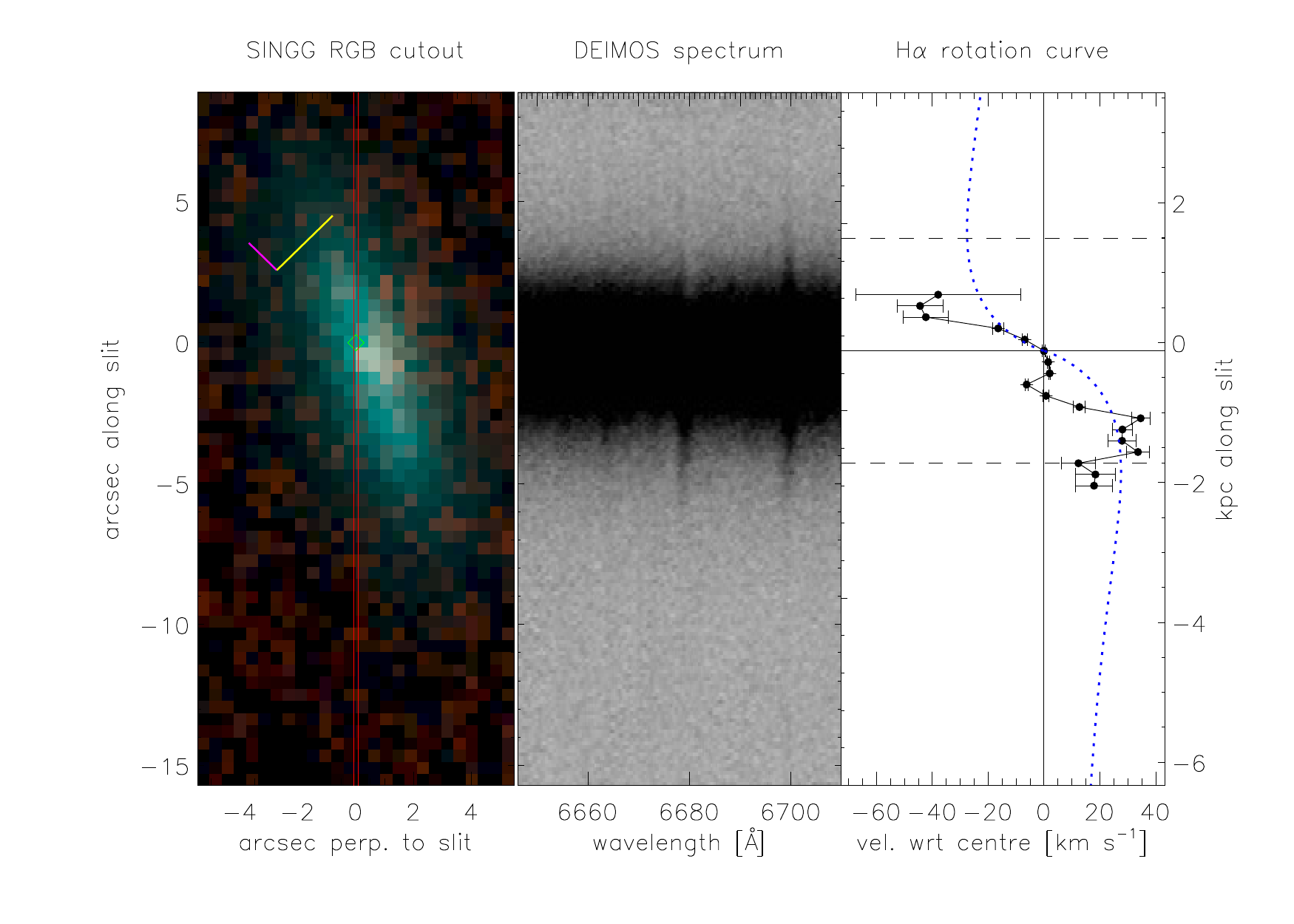}
}
\caption{J1051-17:S8. As for Figure~\ref{0443_S4}. 
\label{1051_S8}}
\end{figure*}

{ J1051-17:S8} (Figure~\ref{1051_S8}). The velocity profile of this galaxy is consistent with a mass-follows-light profile with a kinematically decoupled core. The faint H$\alpha$ emission and bright continuum at the centre makes the velocity profile difficult to measure beyond the region shown here, especially at the NE end of the slit. The apparent H$\alpha$ absorption in the central panel is in fact an adjacent, poorly-subtracted sky line.

\begin{figure*}
\centerline{
\includegraphics[width=0.75\linewidth]{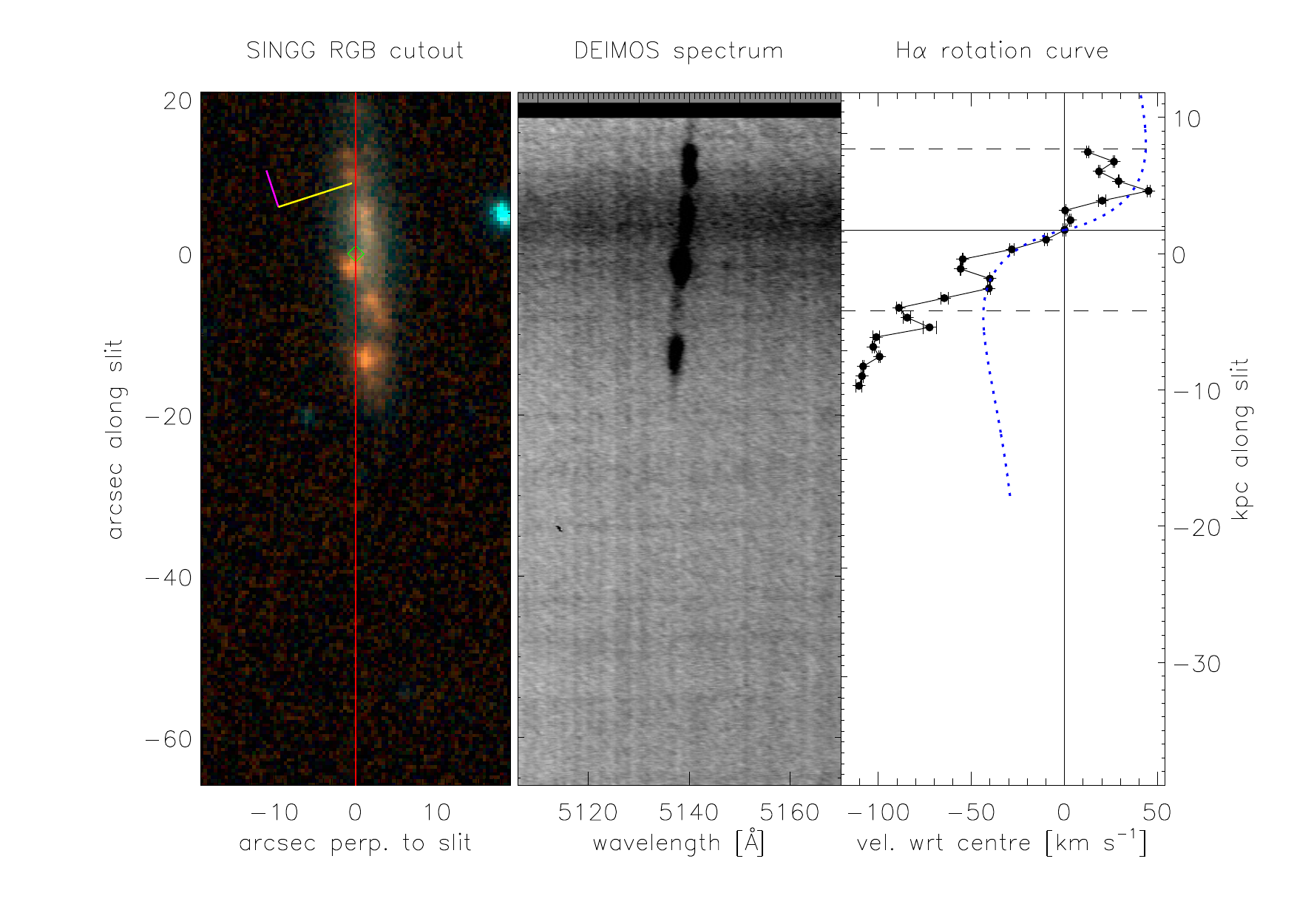}
}
\caption{J1059-09:S5. As for Figure~\ref{0443_S4}. 
\label{1059_S5}}
\end{figure*}

{ J1059-09:S5} (Figure~\ref{1059_S5}). For this slit, H$\alpha$ and [N{\sc ii}] lines fall on the edge of the CCD, so we opt for the [O{\sc iii}] $\lambda$5007 line in our kinematical analysis in order to avoid possible edge of frame effects. Similar observations with a slit better placed with respect to the mask centre would result in H$\alpha$ kinematics extending $\sim$1 kpc beyond what we show for [O{\sc iii}] 5007. The observed velocity rises above the model curves, suggestive of the presence of DM.

\begin{figure*}
\centerline{
\includegraphics[width=0.75\linewidth]{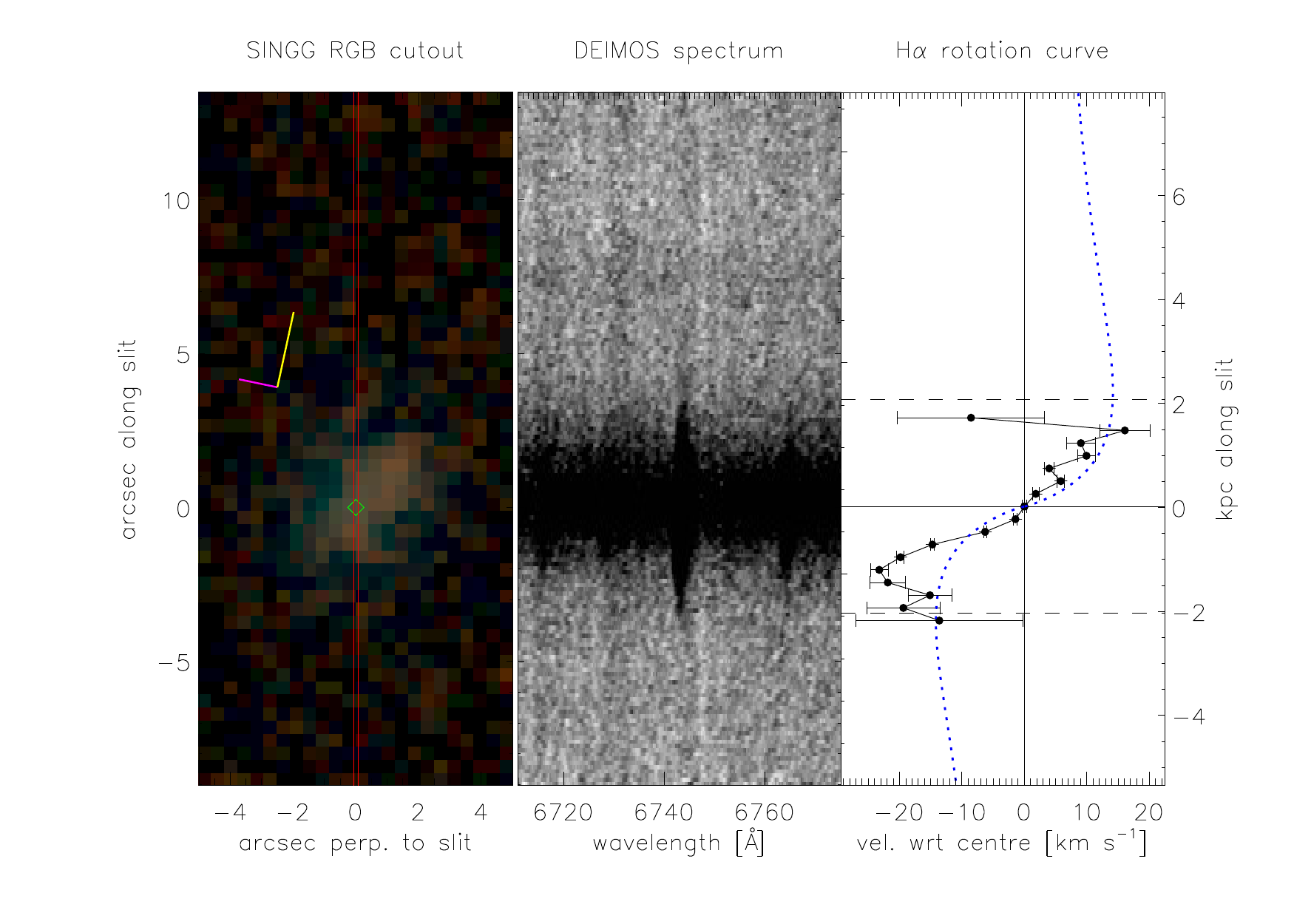}
}
\caption{J1059-09:S8. As for Figure~\ref{0443_S4}. 
\label{1059_S8}}
\end{figure*}

{ J1059-09:S8} (Figure~\ref{1059_S8}). Mass follows light for this small galaxy. It has a metallicity 2.5$\sigma$ above the SDSS mean for its luminosity, so is close to our TDG selection limit. Moreover, it is near in projection to the very high metallicity dwarf S10 and may be related to it.

\begin{figure*}
\centerline{
\includegraphics[width=0.75\linewidth]{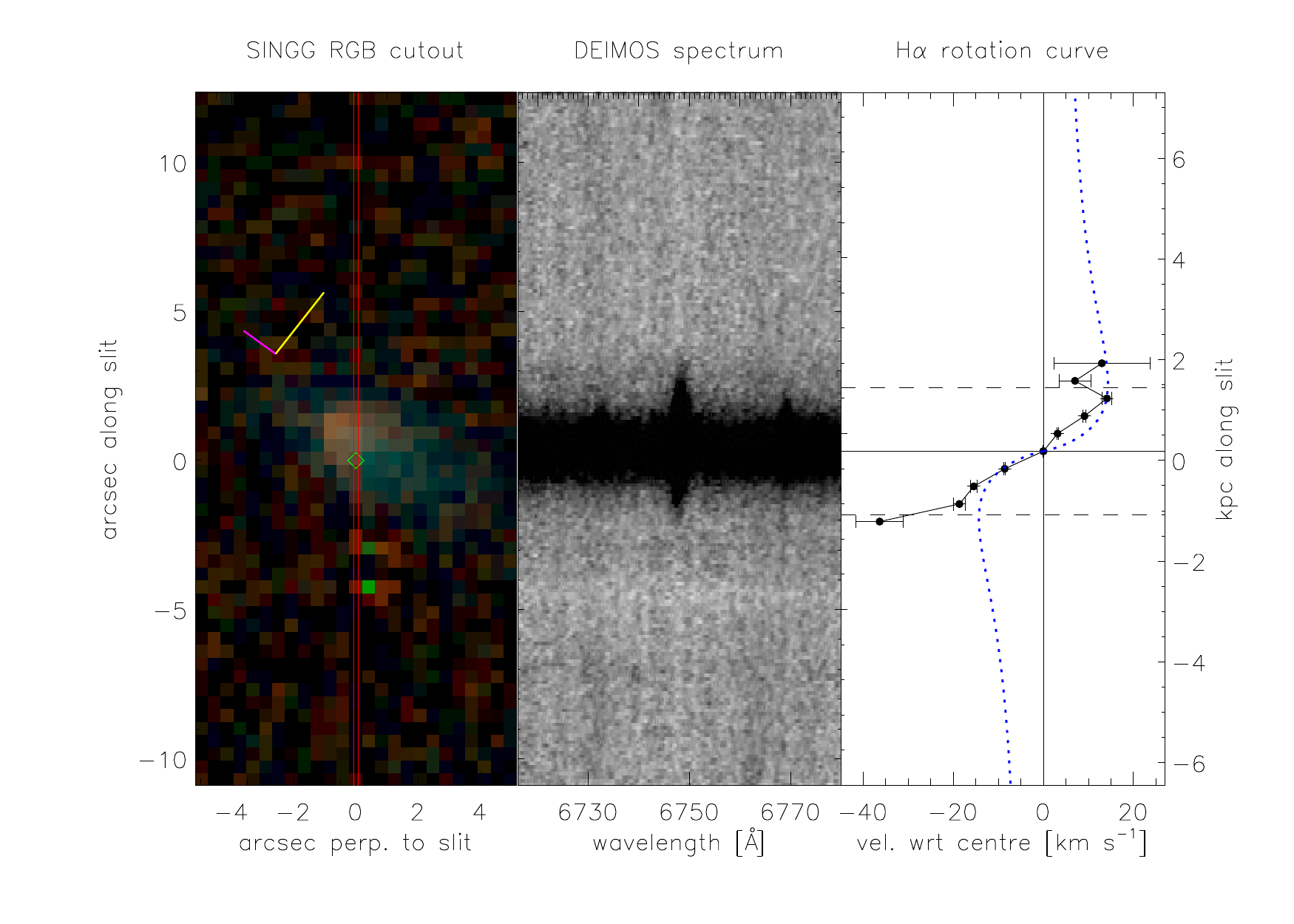}
}
\caption{J1059-09:S9. As for Figure~\ref{0443_S4}.
\label{1059_S9}}
\end{figure*}

{ J1059-09:S9} (Figure~\ref{1059_S9}). This small galaxy has very low surface brightness.

\subsection{Figures}

\clearpage

\section{Sample B}

\subsection{Notes on individual galaxies}

\begin{figure*}
\centerline{
\includegraphics[width=0.75\linewidth]{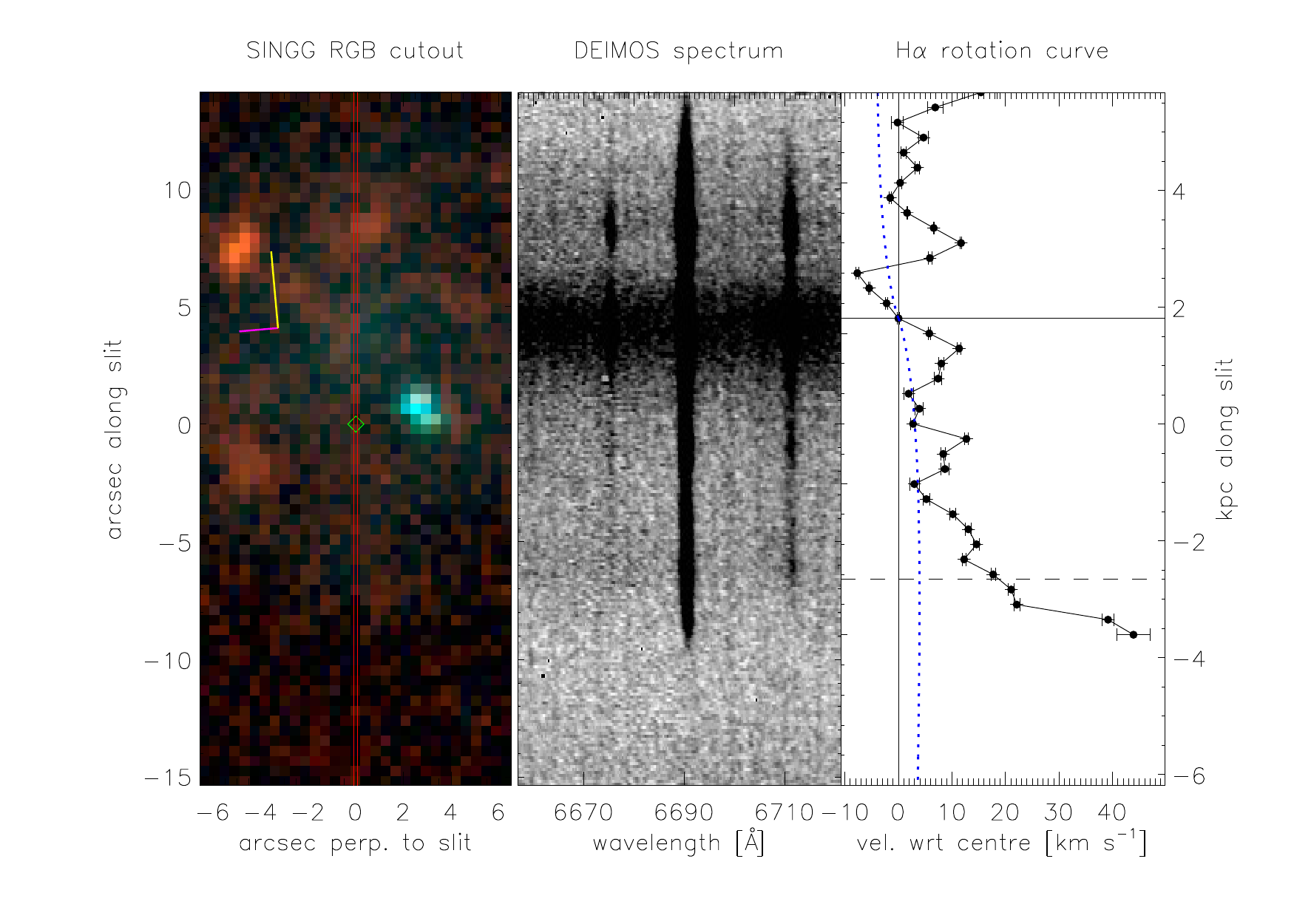}
}
\caption{J1051-17:S3. As for Figure~\ref{0443_S4}. 
\label{1051_S3}}
\end{figure*}

{ J1051-17:S3} (Figure~\ref{1051_S3}). This galaxy is face-on, so no meaningful velocity profile is measurable.

\begin{figure*}
\centerline{
\includegraphics[width=0.75\linewidth]{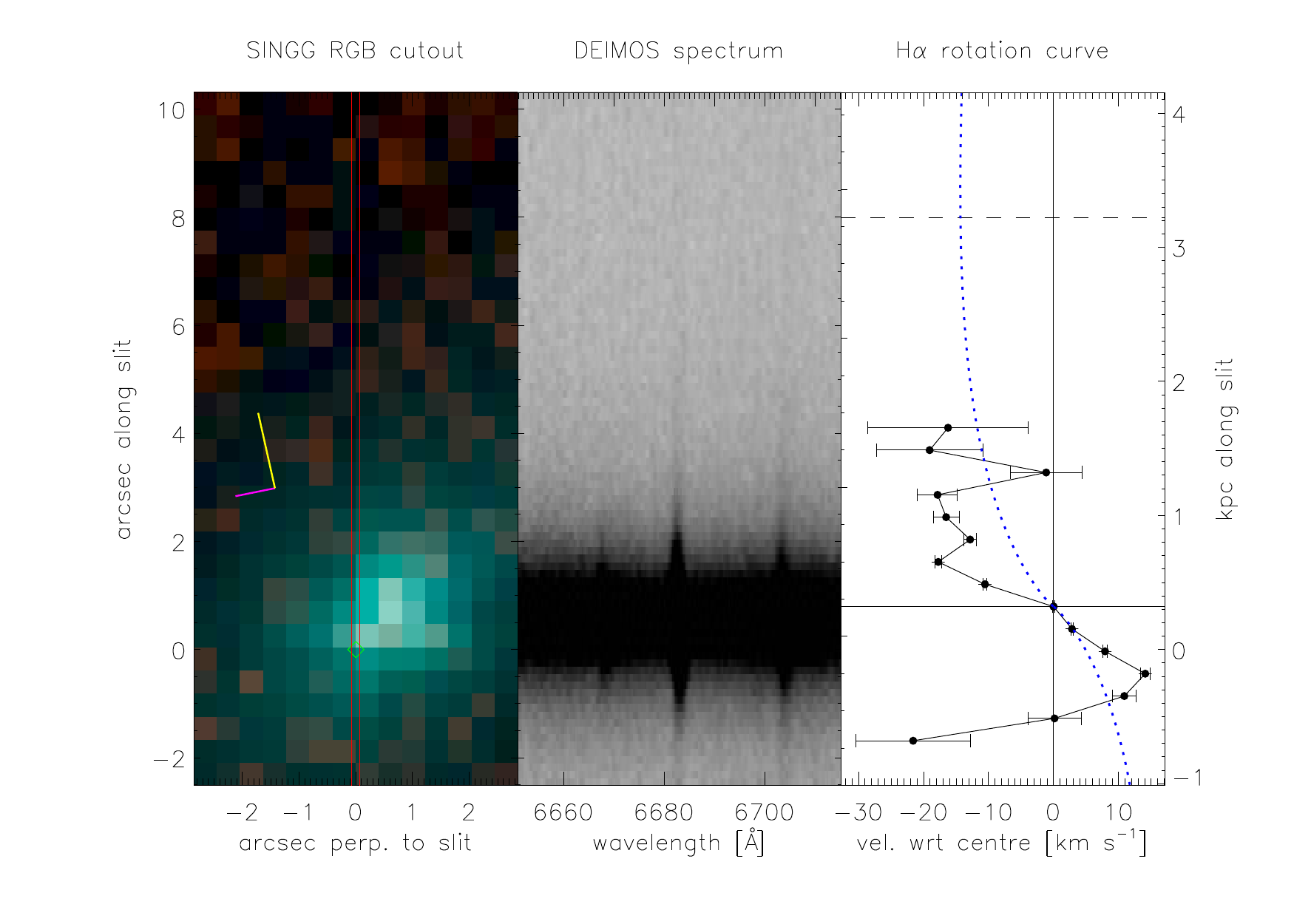}
}
\caption{J1051-17:g04. As for Figure~\ref{0443_S4}. 
\label{1051_g04}}
\end{figure*}

{ J1051-17:g04} (Figure~\ref{1051_g04}). The misalignment between optical and slit PAs prohibits sound measurement of this small, newly-identified galaxy.

\begin{figure*}
\centerline{
\includegraphics[width=0.75\linewidth]{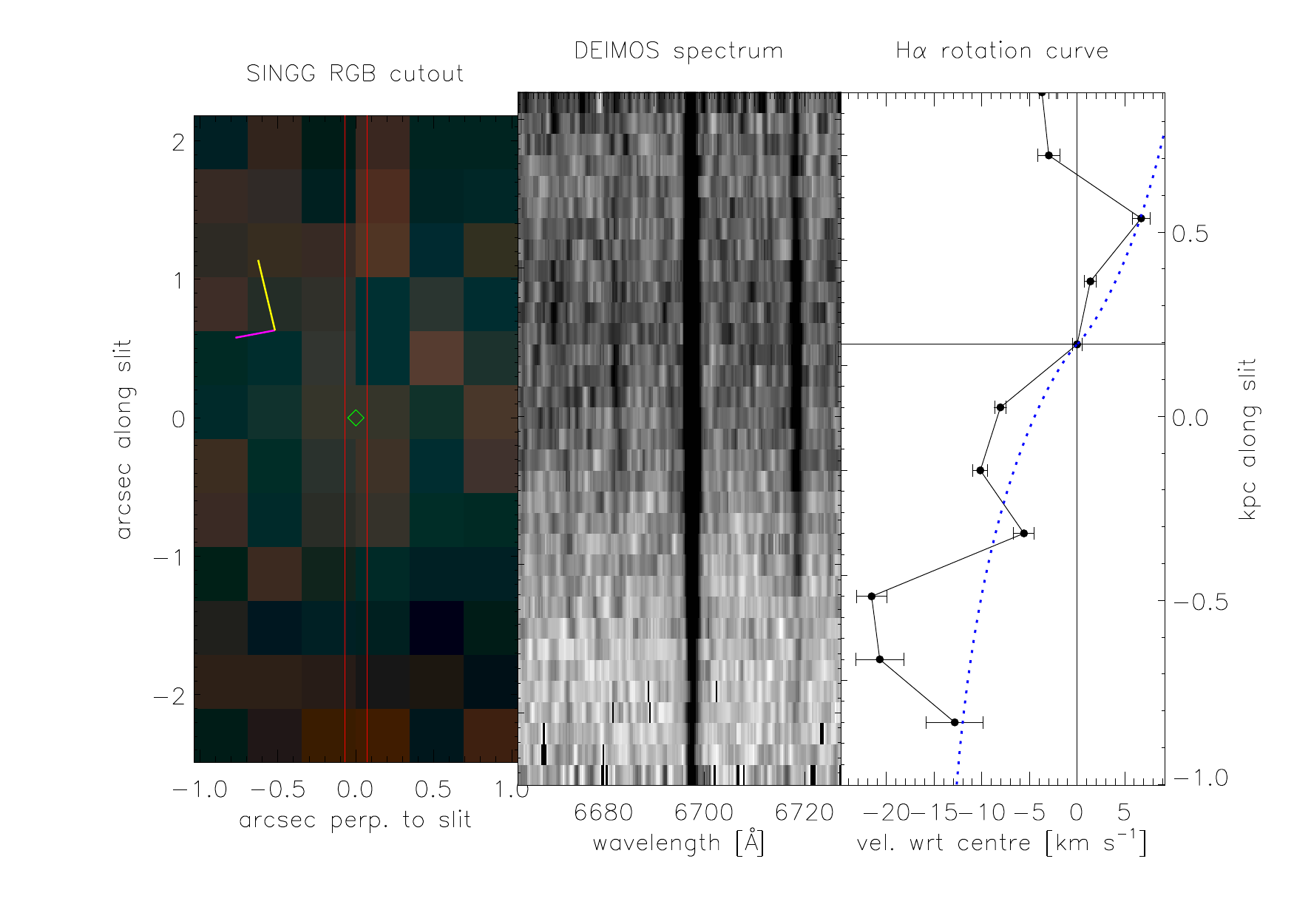}
}
\caption{J1051-17:g07. As for Figure~\ref{0443_S4}. 
\label{1051_g07}}
\end{figure*}

{ J1051-17:g07} (Figure~\ref{1051_g07}). This new, faint galaxy has strong H$\alpha$ emission but very faint continuum.

\begin{figure*}
\centerline{
\includegraphics[width=0.75\linewidth]{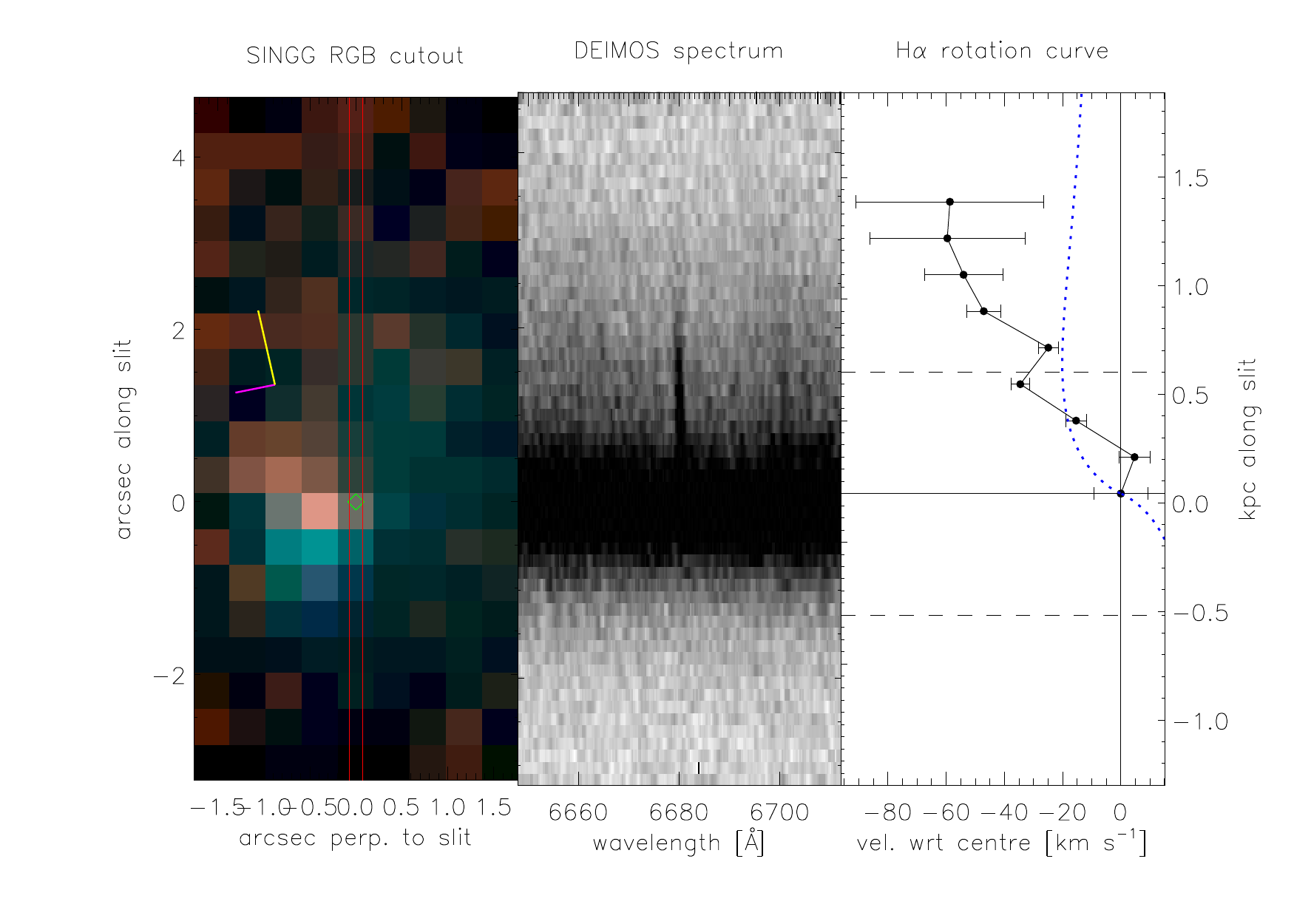}
}
\caption{J1051-17:g11. As for Figure~\ref{0443_S4}. 
\label{1051_g11}}
\end{figure*}

{ J1051-17:g11} (Figure~\ref{1051_g11}). For this new galaxy there is a very large correction for inclination and orientation. There are no observed data points below the continuum centre in Figure~\ref{1051_g11} because there is insufficient H$\alpha$ light on this side of the galaxy. Its high metallicity leads us to classify this galaxy as a very strong TDG candidate.

\begin{figure*}
\centerline{
\includegraphics[width=0.75\linewidth]{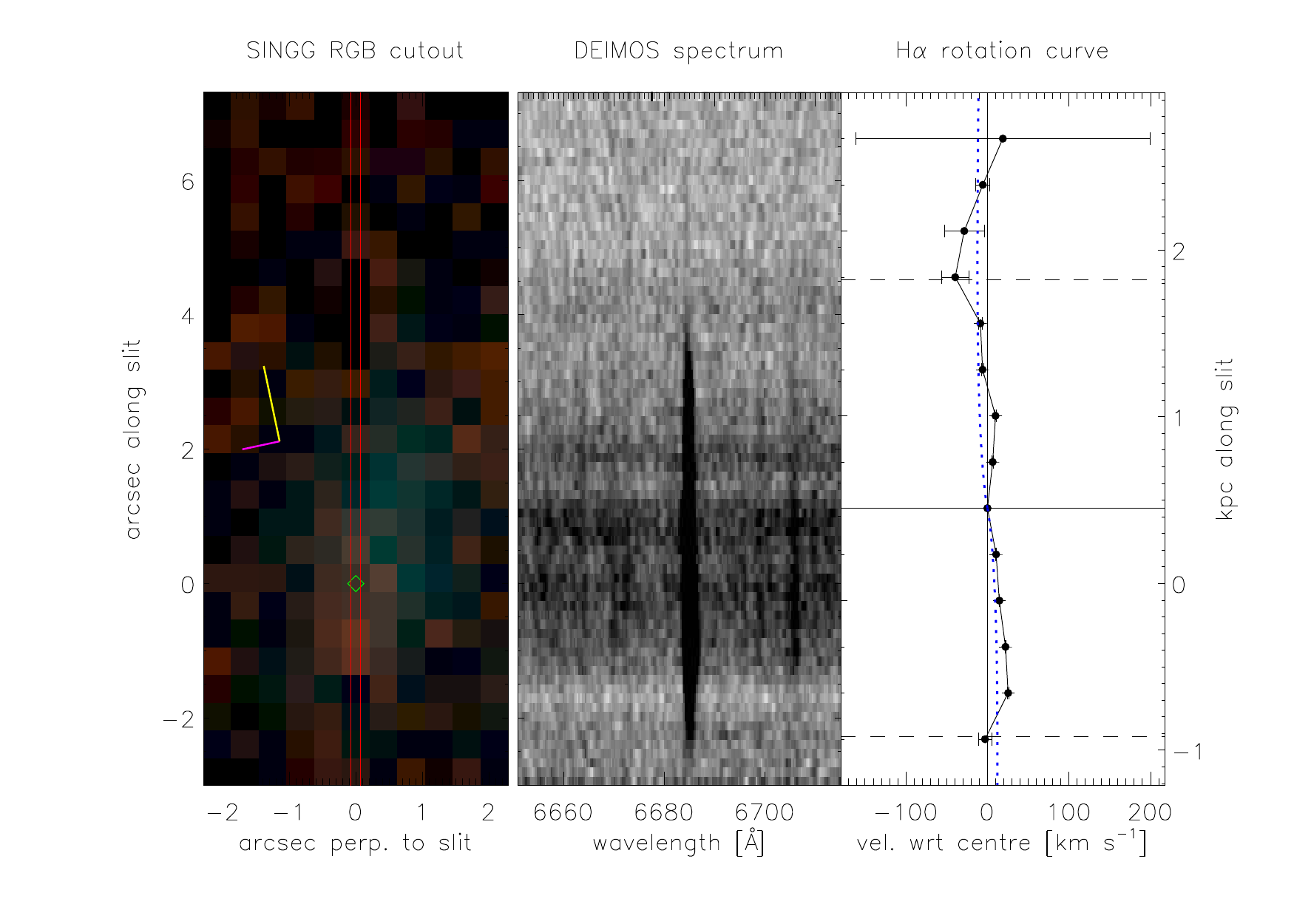}
}
\caption{J1051-17:g13. As for Figure~\ref{0443_S4}. 
\label{1051_g13}}
\end{figure*}

{ J1051-17:g13} (Figure~\ref{1051_g13}). This galaxy's inclination is uncertain, so it is difficult to measure its mass-to-light ratio. Its velocity profile appears to be disturbed.

\begin{figure*}
\centerline{
\includegraphics[width=0.75\linewidth]{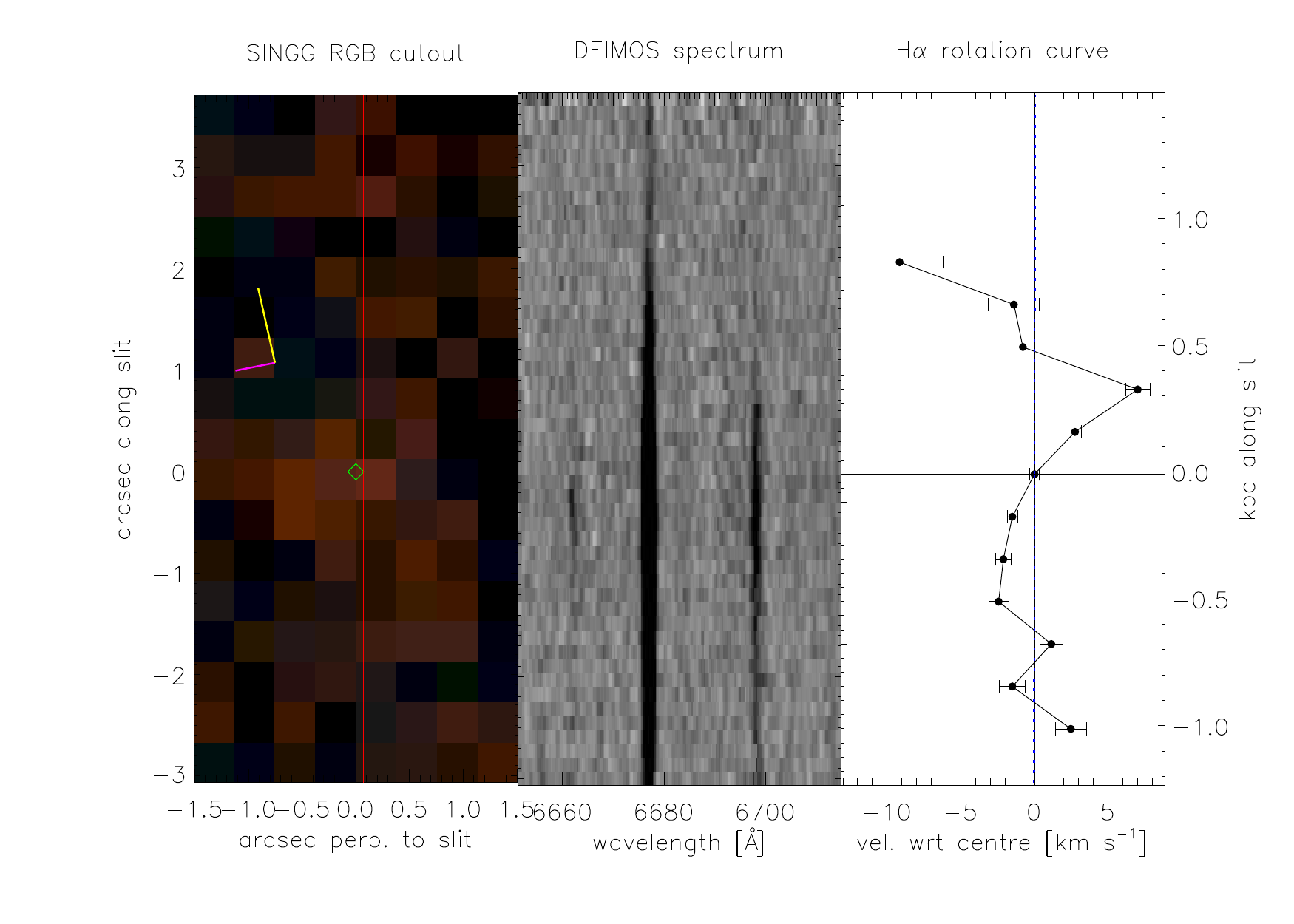}
}
\caption{J1051-17:g15. As for Figure~\ref{0443_S4}. 
\label{1051_g15}}
\end{figure*}

{ J1051-17:g15} (Figure~\ref{1051_g15}). This galaxy has very low surface brightness, so is barely evident in the SINGG image. Its extent in H$\alpha$ is well below the tidal truncation size and the velocity structure along the slit is erratic but has low amplitude. 

\begin{figure*}
\centerline{
\includegraphics[width=0.75\linewidth]{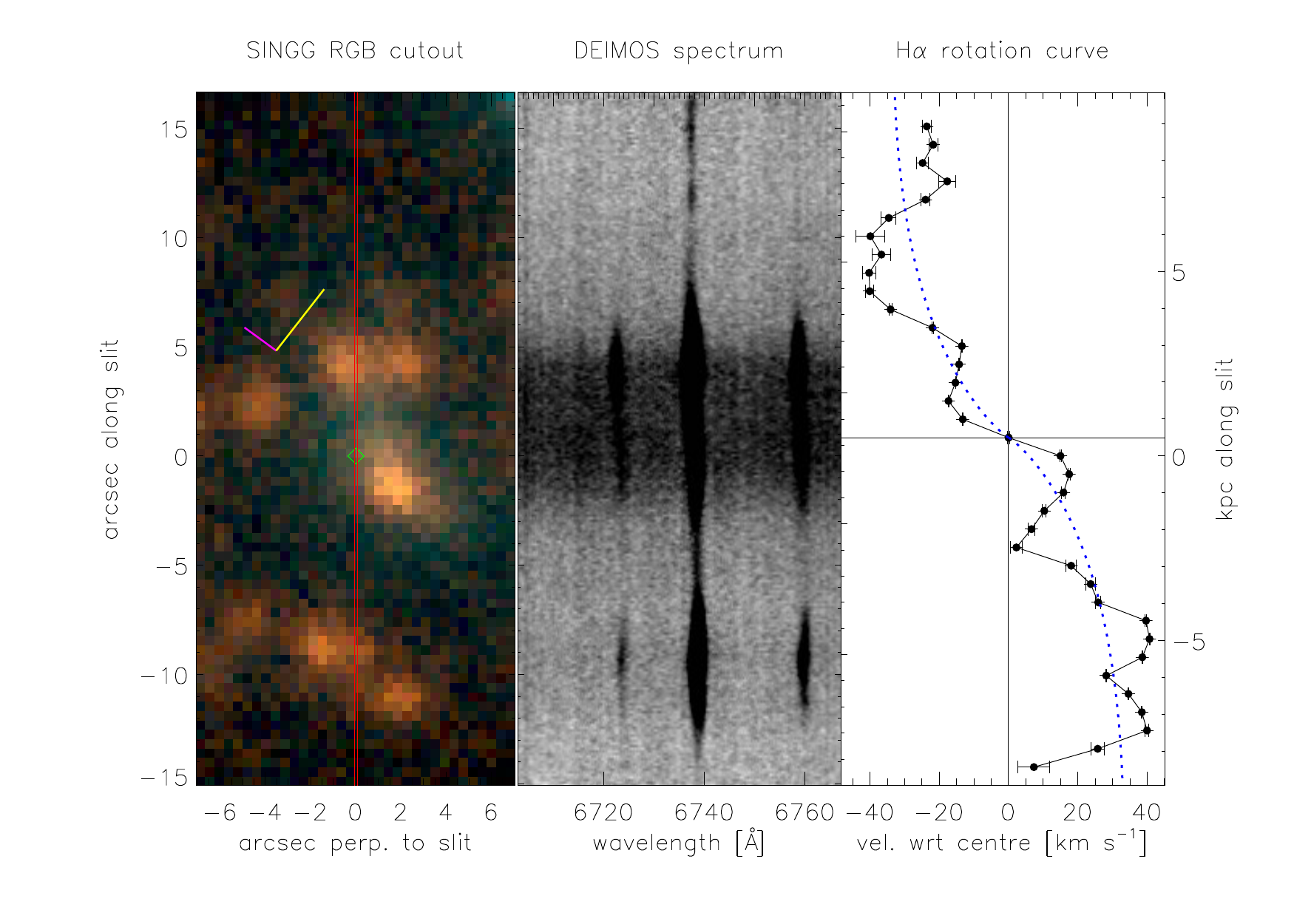}
}
\caption{J1059-09:S2. As for Figure~\ref{0443_S4}. 
\label{1059_S2}}
\end{figure*}

{ J1059-09:S2} (Figure~\ref{1059_S2}). This clumpy galaxy is consistent with a mass-follows-light profile. Note that the large corrections for slit position angle and galaxy inclination contribute to the large M/L ratio.

\begin{figure*}
\centerline{
\includegraphics[width=0.75\linewidth]{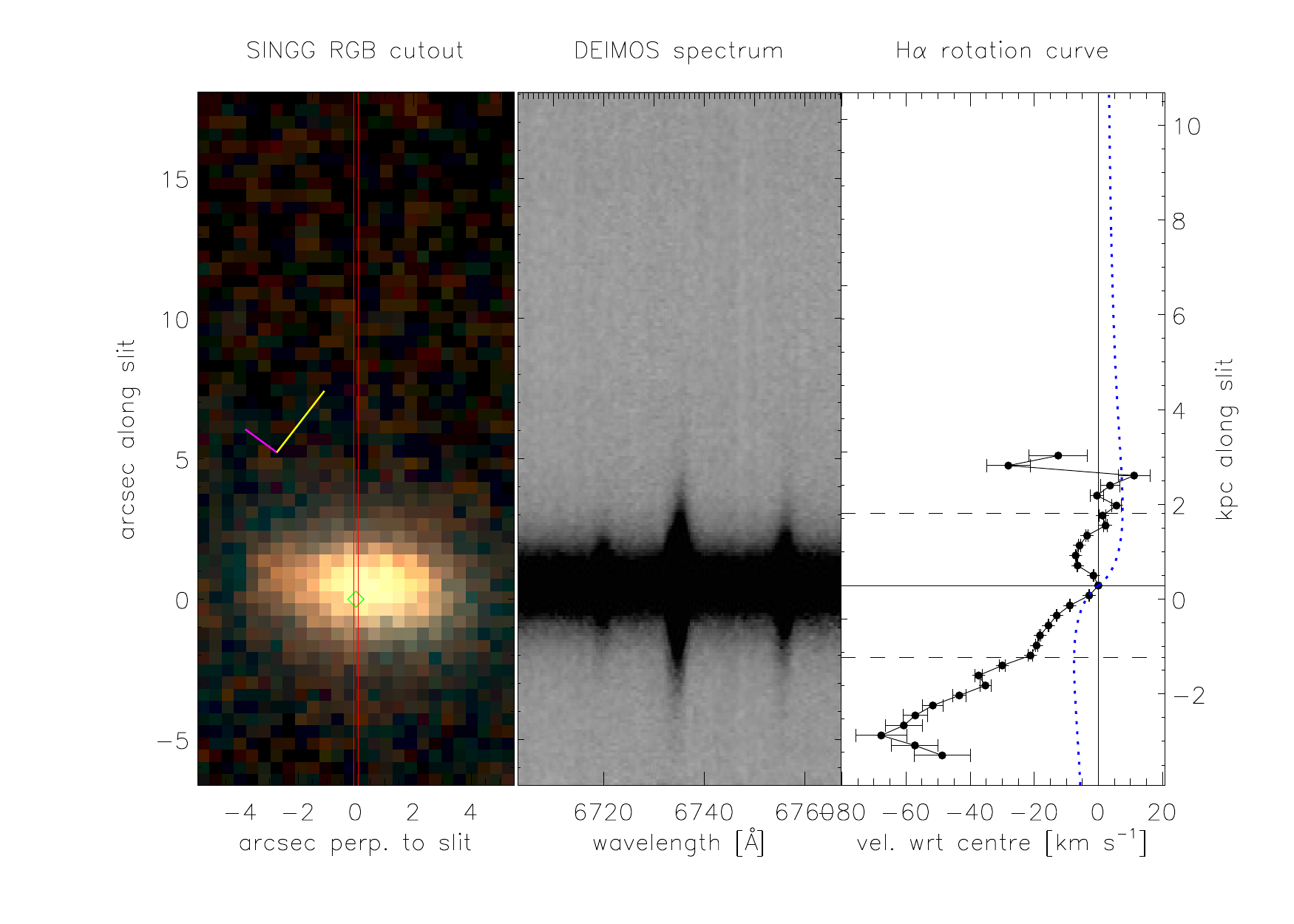}
}
\caption{J1059-09:S7. As for Figure~\ref{0443_S4}. 
\label{1059_S7}}
\end{figure*}

{ J1059-09:S7} (Figure~\ref{1059_S7}). The velocity profile of this galaxy has a very unusual shape. It is very near the interacting galaxies S1 and S3, so is likely disturbed by their tides. Moreover, the slit is close to aligning with the minor axis of this galaxy, so the kinematics may be indicative of a galactic wind.

\begin{figure*}
\centerline{
\includegraphics[width=0.75\linewidth]{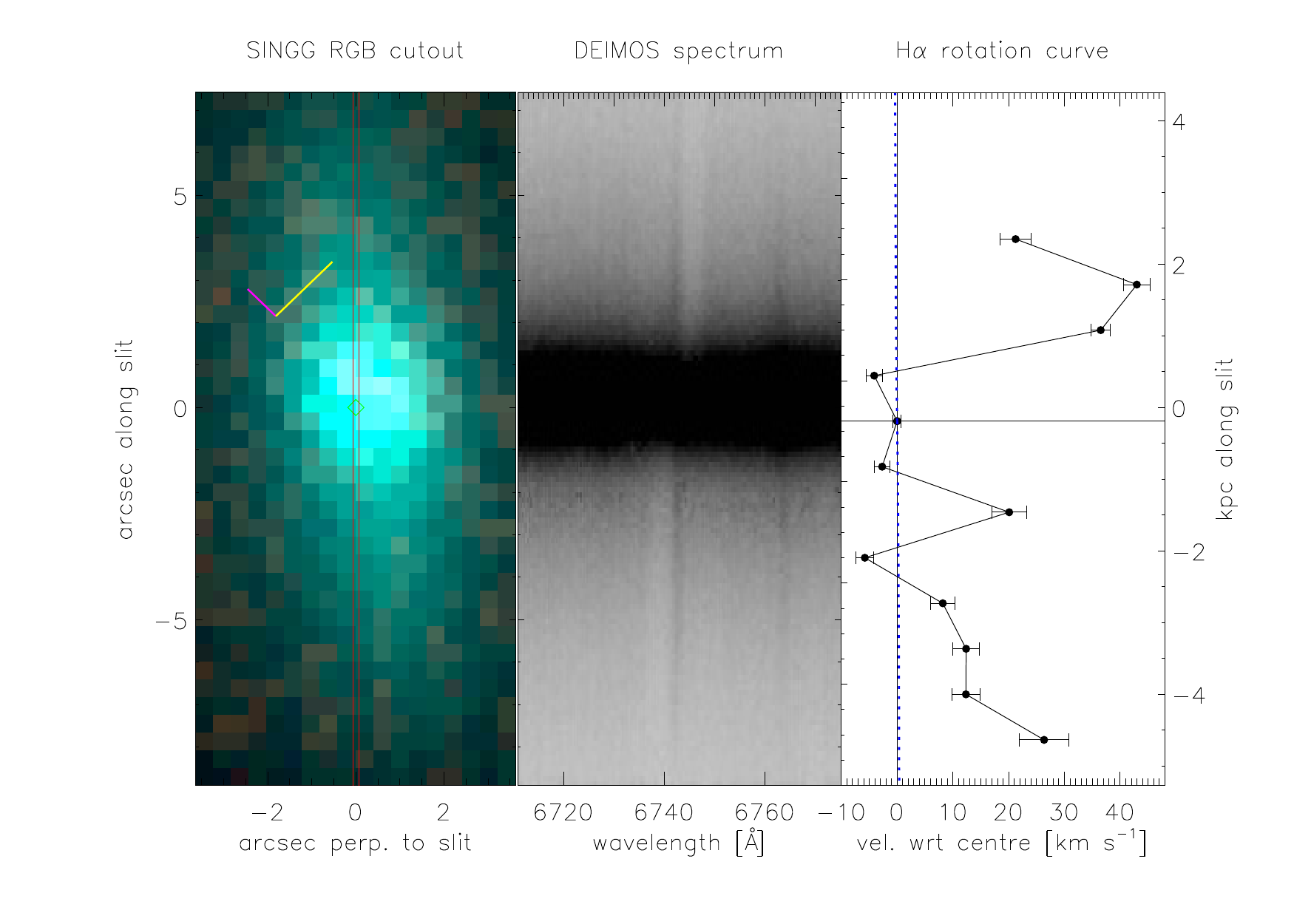}
}
\caption{J1059-09:S10. As for Figure~\ref{0443_S4}. 
\label{1059_S10}}
\end{figure*}

{ J1059-09:S10} (Figure~\ref{1059_S10}). This galaxy has a very strong stellar component (as evidenced by the blue colour in Figure~\ref{1059_S10}). Its rotation curve is not well fit by a predicted mass-follows-light relation. The galaxy has a possible counter-rotating H$\alpha$ absorption component, the analysis of which is beyond the scope of the paper. It has a fairly high metallicity consistent with a TDG, though not above our diagnostic cut. However, it is not near a host, so must be an old TDG if it is one. It is near another dwarf (S8) of metallicity $\sim$2.5-$\sigma$ above the SDSS control sample.

\begin{figure*}
\centerline{
\includegraphics[width=0.75\linewidth]{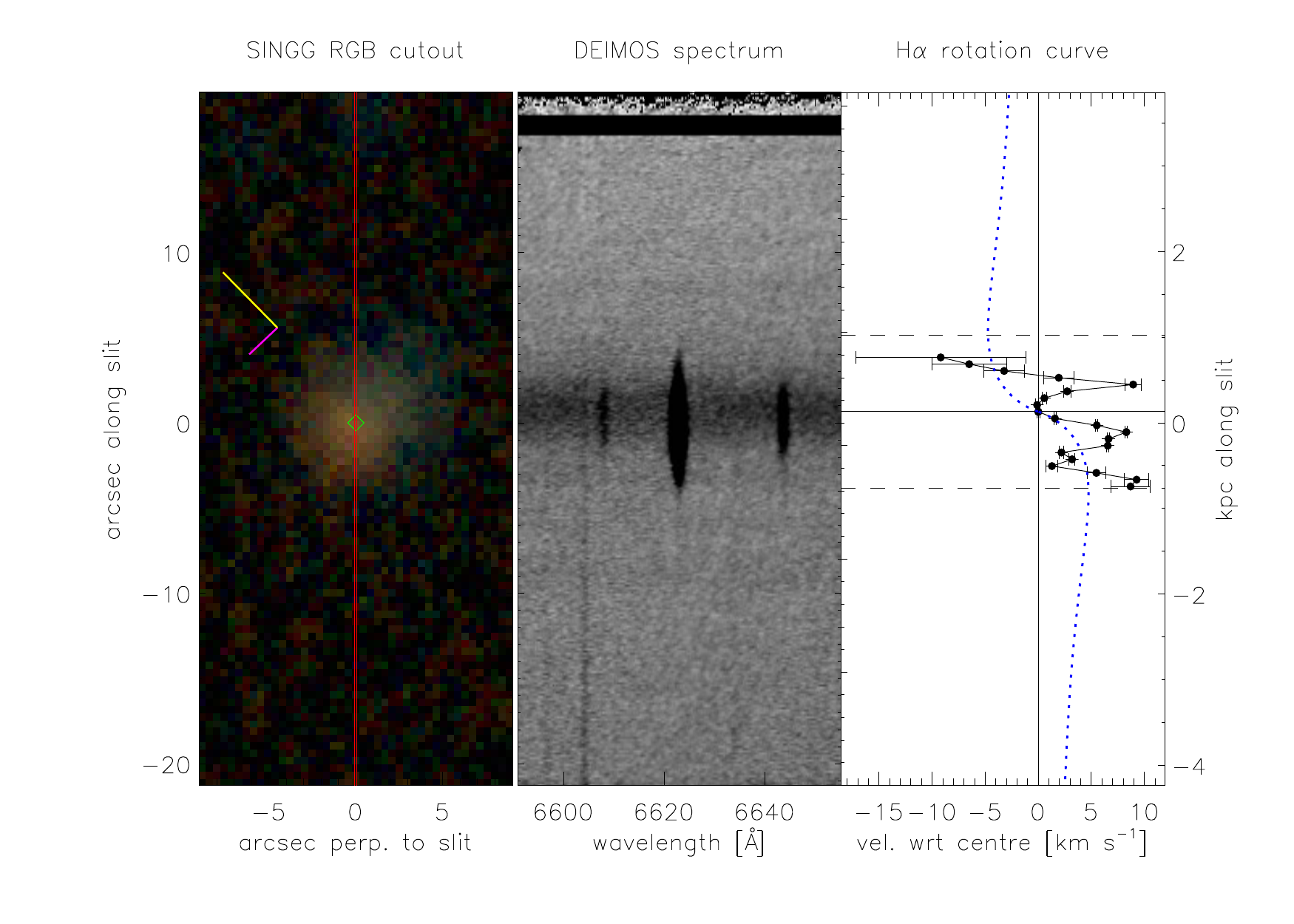}
}
\caption{J1403-06:S3. As for Figure~\ref{0443_S4}.  
\label{1403_S3}}
\end{figure*}

{ J1403-06:S3} (Figure~\ref{1403_S3}). This slit measures an offset HII region within a very low surface brightness dwarf galaxy, so it is difficult to claim that this galaxy is rotating. It is only marginally detectable based on our $r_{tidal}$ detection limit, and indeed we do not measure rotation past $r_{turn}$.

\begin{figure*}
\centerline{
\includegraphics[width=0.75\linewidth]{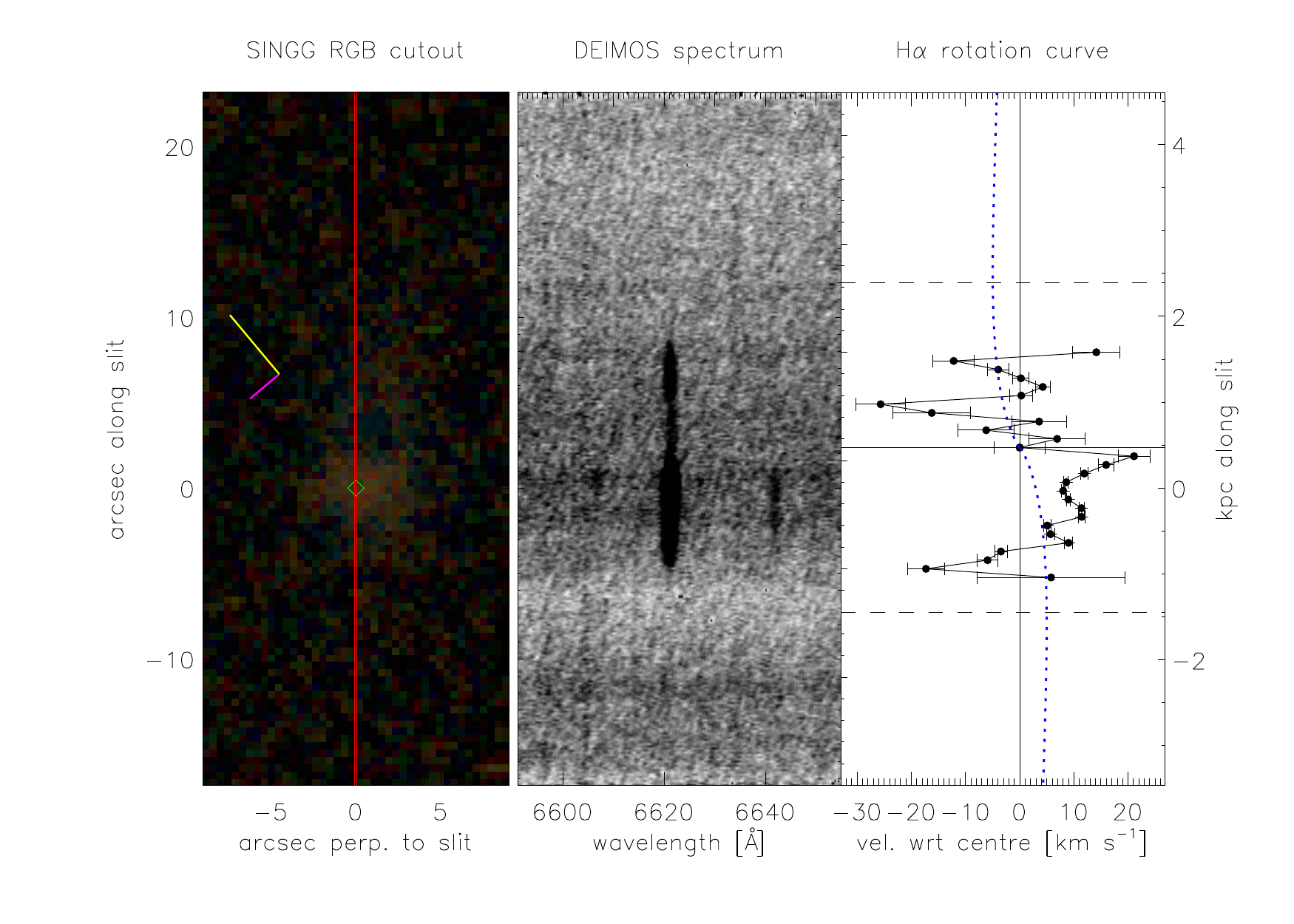}
}
\caption{J1403-06:S4. As for Figure~\ref{0443_S4}.
\label{1403_S4}}
\end{figure*}

{ J1403-06:S4} (Figure~\ref{1403_S4}). This very low surface brightness galaxy consists of two SF regions likely disturbed by the two giant interacting galaxies S1 and S2. It is sufficiently small to lie below the detection limit set by $r_{tidal}$.

\begin{figure*}
\centerline{
\includegraphics[width=0.75\linewidth]{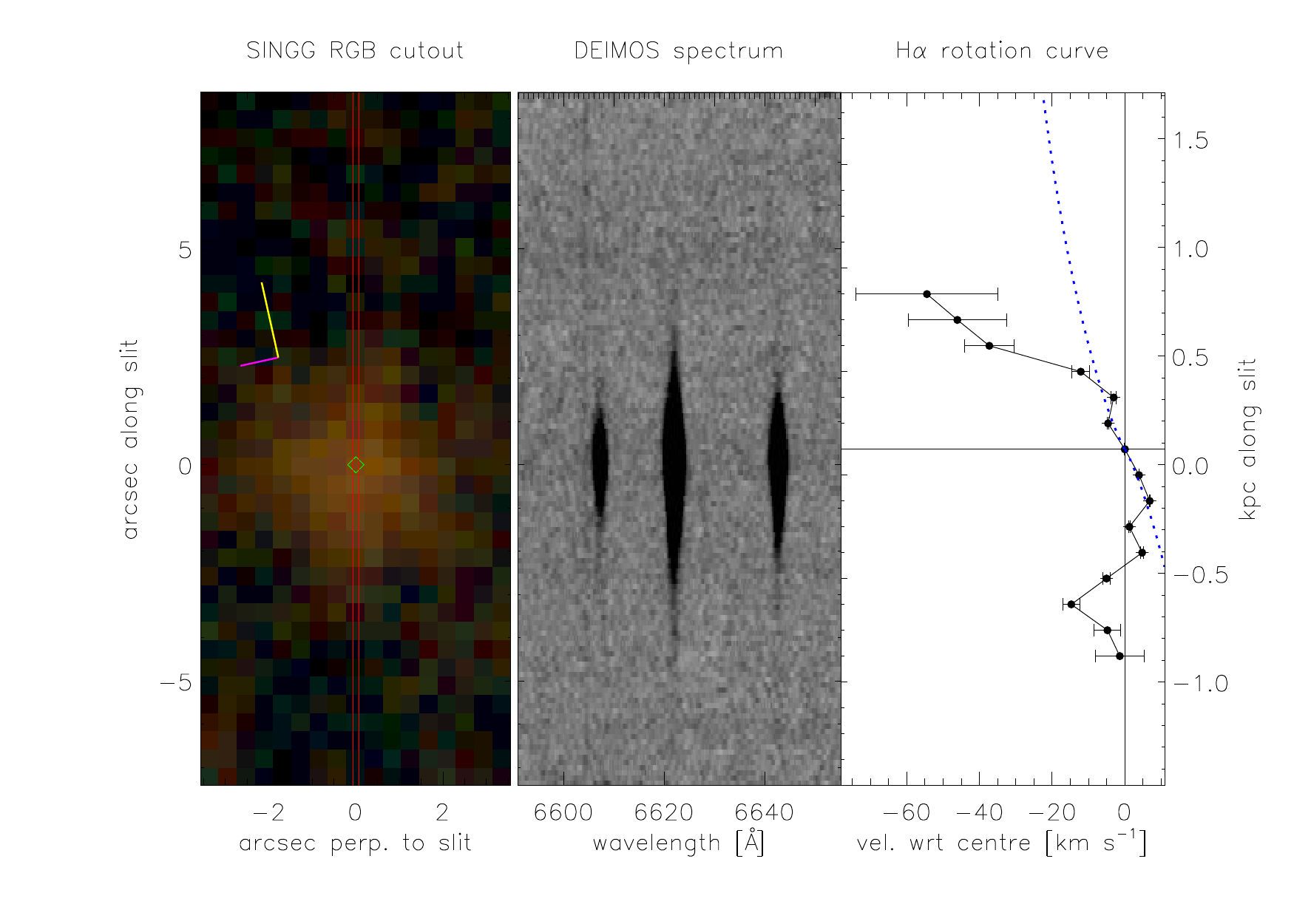}
}
\caption{J1403-06:g1. As for Figure~\ref{0443_S4}. 
\label{1403_g1}}
\end{figure*}

{ J1403-06:g1} (Figure~\ref{1403_g1}). This appears to be a giant HII region within the halo of the giant galaxy S2. It has high metallicity consistent with a TDG, but a very warped rotation curve. Due to its small size and proximity to S2, it is well below the tidal stripping detection limit. 

\subsection{Figures}

\clearpage

\section{Mask and slit placement}

\begin{figure}
\centerline{
\includegraphics[width=0.75\linewidth]{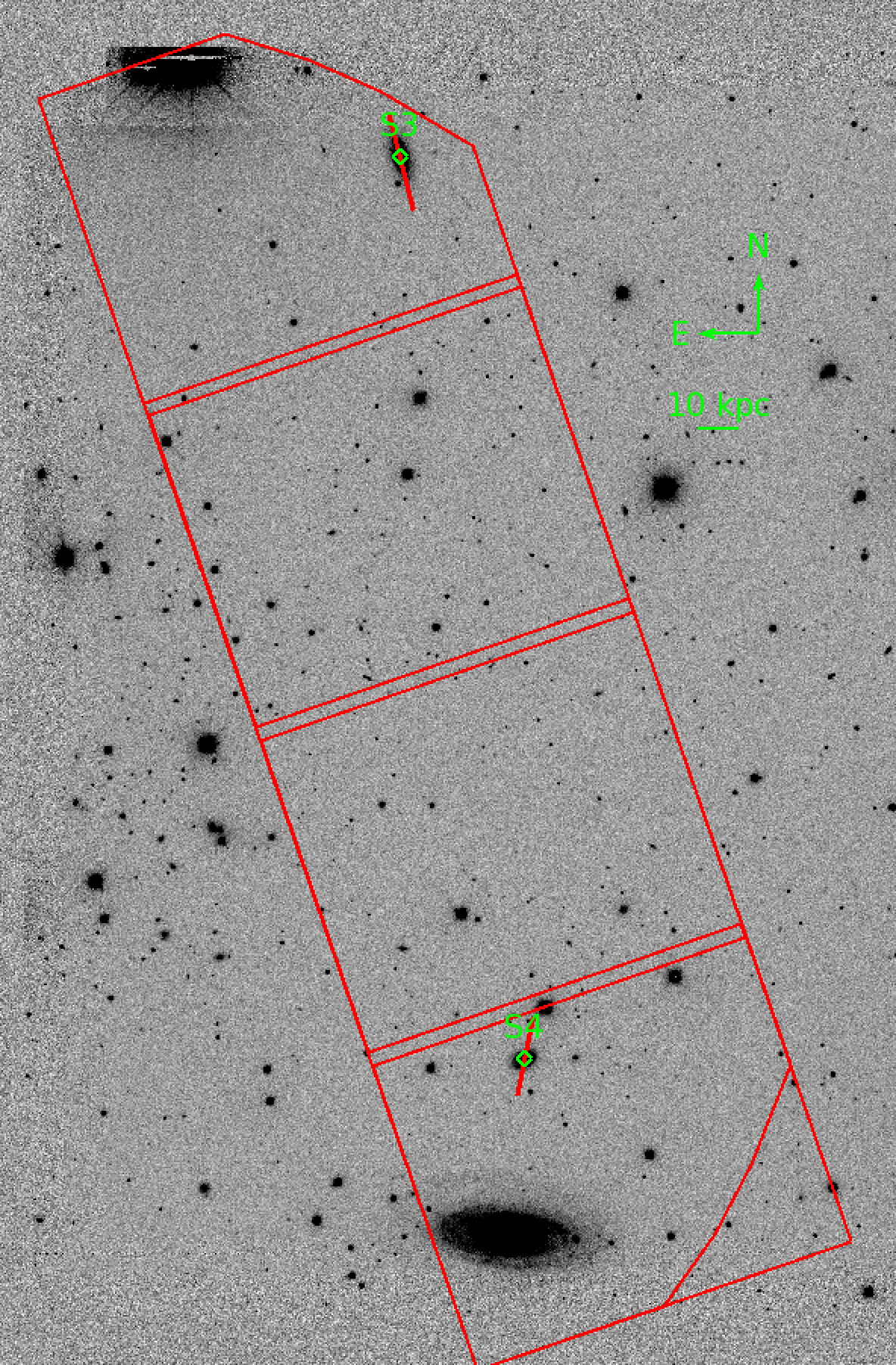}
}
\caption{SINGG image of HIPASS J0443-05 showing locations of slits (red) for measured galaxies (green labels). The approximate location of the DEIMOS mask is given by the red polygon. The other slits are not shown for clarity.
\label{0443}}
\end{figure}

\begin{figure}
\centerline{
\includegraphics[width=0.75\linewidth]{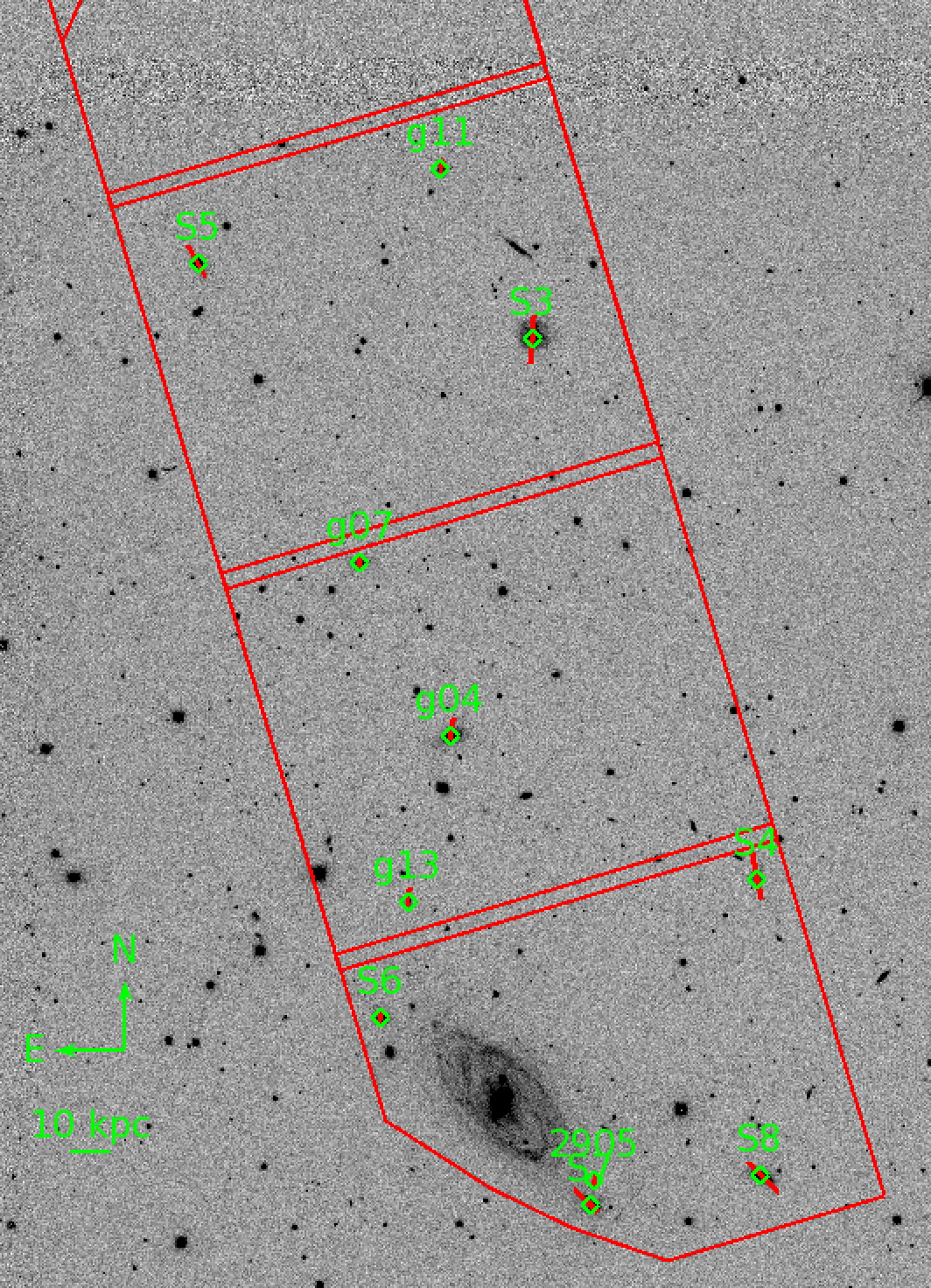}
}
\caption{SINGG image of HIPASS J1051-17 showing locations of slits (red) for measured galaxies (green labels). The approximate location of the DEIMOS mask is given by the red polygon. The other slits are not shown for clarity.
\label{1051}}
\end{figure}

\begin{figure}
\centerline{
\includegraphics[width=0.75\linewidth]{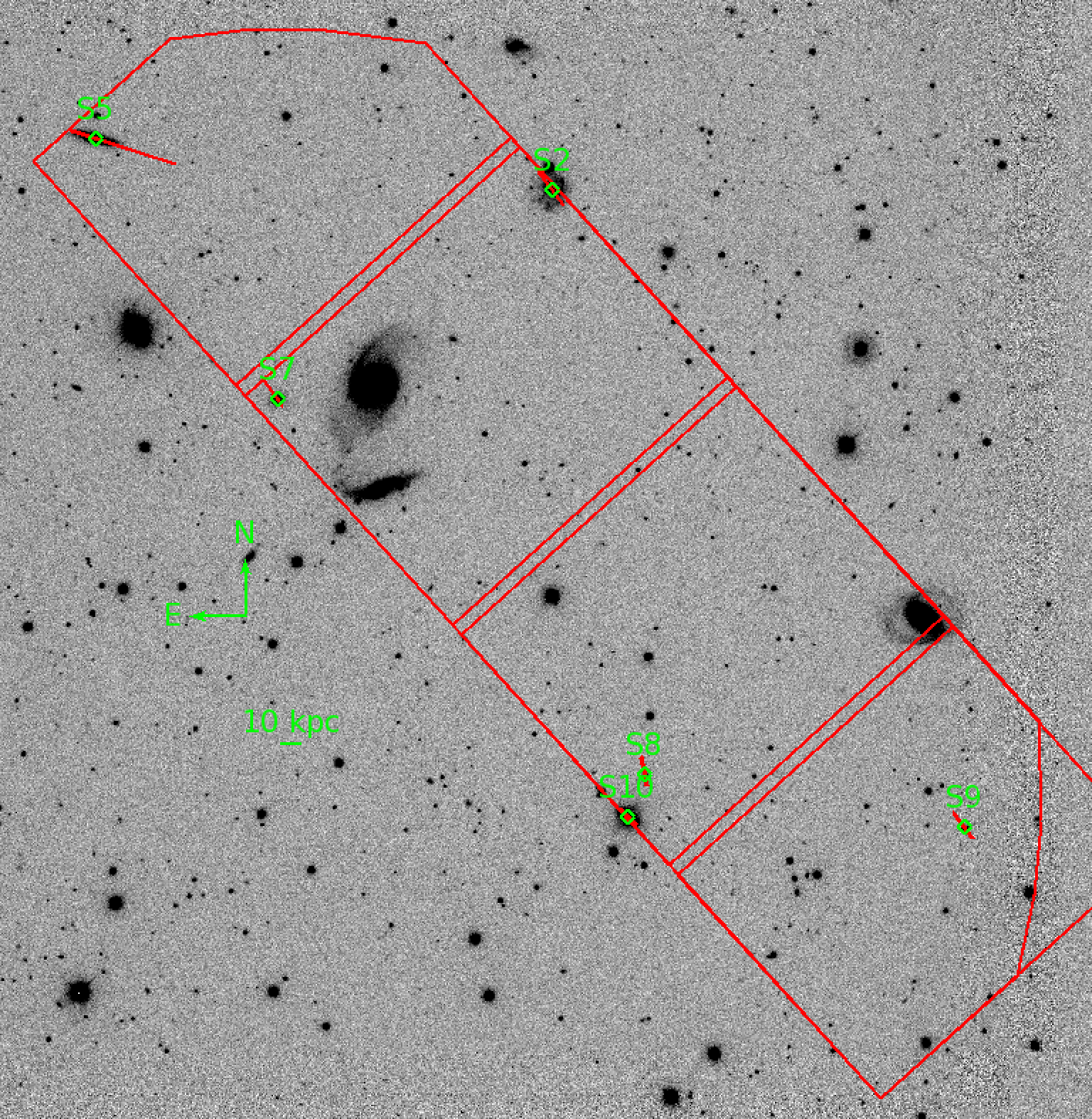}
}
\caption{SINGG image of HIPASS J1059-09 showing locations of slits (red) for measured galaxies (green labels). The approximate location of the DEIMOS mask is given by the red polygon. The other slits are not shown for clarity.
\label{1059}}
\end{figure}

\begin{figure}
\centerline{
\includegraphics[width=0.75\linewidth]{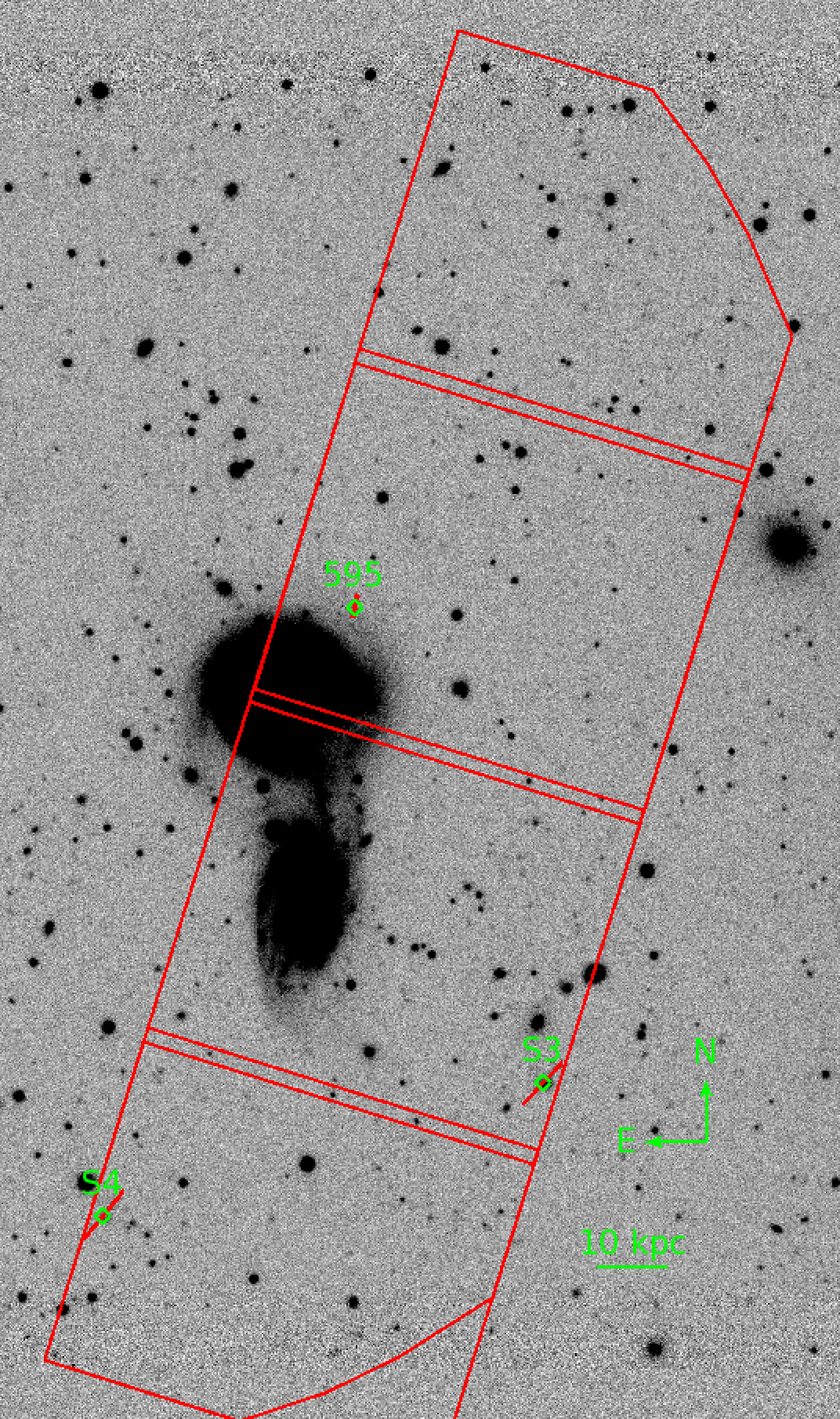}
}
\caption{SINGG image of HIPASS J1403-06 showing locations of slits (red) for measured galaxies (green labels). The approximate location of the DEIMOS mask is given by the red polygon. The other slits are not shown for clarity.
\label{1403}}
\end{figure}

\end{document}